\documentclass[preprint,aps]{revtex4-1}
\usepackage{graphicx}
\usepackage{rotating}

\begin{document}
\title{Extensive Nature of Long Range Interactions: Role of Disorder}
\author{Pragya Shukla}
\affiliation{Department of Physics, Indian Institute of Technology, Kharagpur-721032, India}
\date{\today}

\begin{abstract}

The omnipresent disorder in physical systems  makes it imperative to investigate its effect on the  spatial range of interactions for which system remains thermodynamically extensive.
Previously known  bounds   on the statistical free energy for clean systems \cite{fish} indicate it to be extensive only for the  spatially short range interactions (decaying faster than  $r^{-d}$ at large distance $r$ with $d$ as system dimension).    We analyze the bounds for quantum systems with different types of disordered many body potentials e.g annealed, quenched, Gaussian or power law distributed. Our results indicate the dependence of the bounds on  the multiple distribution  parameters representing the potential which  in turn permits, in contrast to clean potentials, more freedom to achieve the  extensive limits even for arbitrary spatial ranges of the interactions.

\end{abstract}


\maketitle

       .

{

\section{Introduction}

The complexity in varied forms e.g.  many-body interactions, disorder etc. in real physical systems makes it necessary to explore their statistical properties  and the approach to  thermodynamic limits. 
An important role in this context is played by the range of many body interactions (the spatial decay of  interaction at large distances relative to its dimensionality). Based on the range, the interaction can be  classified in two categories: (i) short range interactions  (SRI) that fall off faster than $r^{-\alpha}$  for sufficiently large distances $r$ between particle-pairs with $\alpha > d$, and (ii) long-range interactions (LRI) with $\alpha \le d$ with $d$ as space-dimension. (It must be noted that the ''range'' mentioned here is different from the characteristic length-scale  of the  potential).  The peculiar thermodynamic behavior of LRIs  has motivated many studies in recent past \cite{lri1, lri2, lri3} and is also  primary focus of the present work. The specific issue we  address here is regarding the influence of disorder on extensive behavior of quantum systems with long-range interactions. The idea to pursue this study originates from  intense current interest in the questions of localization, thermalization and ergodicity, at finite temperature, of a many body system, isolated or weakly coupled to a bath \cite{nand1, nand2, deut, sred}.

In thermodynamics, an extensive property of a macroscopic system e.g  free energy is defined to be directly proportional to the size of the system and  independent of its shape. In statistical mechanics context however the definition   is not so straight forward. For example, the free energy of a finite system defined through the partition function is not, in general directly proportional to its volume and is shape-dependent. Due to  varying 
definition of partition function across thermodynamics ensembles,  the latter's choice also play an important role. To reconcile the thermodynamics with 
statistical mechanics,  it is therefore necessary that the statistical properties should approach their thermodynamic behavior in the "thermodynamic limit",  
that is,  the limit of infinitely large system-size while keeping the particle density finite. 

As indicated by previous studies \cite{rue,fish,grif, bgw,cdr},  the existence of a "thermodynamic limit"  in a clean system depends on the nature of the interaction which in general may have both attractive as well repulsive parts. This is intuitively expected: an unconstrained increase in the attractive forces in macroscopic limit would lead to collapse of the system, with free energy per particle diverging to $-\infty$. Similarly an unrestricted increase in the repulsive part with increasing volume would cause system to disintegrate with free energy per particle diverging to $\infty$.   The stability of the system in macroscopic limit is therefore feasible only under certain conditions on the interactions. The  necessary, unavoidable role of statistical mechanics in the analysis of many-body systems has motivated many studies in past to probe these conditions. The most rigorous results have been derived by the studies \cite{rue,fish} on clean systems,  quantum as well classical and under various general conditions on the attractive and repulsive part of the potential energy, on the shapes of the domains confining the system and for canonical as well as grand canonical ensemble. Based on these studies, the crucial role played by the range of  many body interactions in absence of disorder, in context of the  system-stability is now well-understood.

          A real many body system  always contains some disorder.   It is therefore natural to wonder about the role of disorder in presence of many body interactions e.g. how  the disorder would affect the allowed "range" of interaction in context of extensive behavior of the physical properties. The intuition suggests that the  disorder may act as a 
barrier (screen) for interaction between two faraway units of the system, thus effectively reducing the "range" of interaction by local-averaging although spatial dependence of the potential (for a single system) may still behave as $r^{-\alpha}$ with $\alpha \le d$. The intuition is indeed supported by the studies in classical long-range lattice models \cite{ks, fz, eh, z, e}) 
but its validity in general 
for classical and quantum systems is not known so far.  As averaging of the properties is necessary for any theoretical/ experimental  comprehension of the disordered systems, the information about effective reduction of the "interaction range" due to disorder, thus increasing its thermodynamic viability,  is very desirable.  This motivates us to reconsider the derivations, given in \cite{fish} of the upper and lower bounds of the free energy for a disordered many body system and seek whether the "range" of interaction can indeed be affected.

Based on underlying complexity e.g. many body interactions, impurities and  scattering conditions etc, the randomness in the system can manifest in various forms which in turn can have significant consequences for  the statistical averages.  In case of the system with an annealed disorder, the random variables it depends on evolve with time; the statistical averages are therefore  carried out over all possible values that the random variables can take. On the contrary, a system with  quenched disorder depends on random variables frozen in time; the  averages are  therefore obtained by keeping the random variables fixed.  The averages also  depend   on the distribution parameters representing the disordered potential as well as on its spatial dependence and a competition among them  is expected to influence the bounds for extensive behavior. This motivates us to consider the disordered potential of both annealed and quenched types, and with distributions of both finite and infinite variances.  Our results, illustrated in Tables I and II for specific cases, clearly indicate the significant role played by disorder to sustain the thermodynamic extensive limits in physical systems: the appearance of multiple distribution parameters in the bounds  indeed helps, by a subtle conspiracy, to  overcome the effect of the spatial range of interactions. In case of annealed disorder, the temperature also appears as a parameter, thus increasing the degree of freedom for the system to approach the thermodynamic limit.

 For clear presentation  of our ideas, here we confine ourselves to disordered potentials in quantum systems in contact with a heat bath which permits the use of canonical ensemble. Note most studies of the LRIs in past have focused on isolated classical systems and therefore analyzed thermodynamic properties in micro-canonical ensemble. The contact of a real disordered LRI with external environment however can not usually  be ignored which makes it necessary to consider canonical ensemble for their analysis.  Our approach  can also be generalized to   grand canonical ensemble along the same lines as discussed in \cite{fish} for clean cases.

The paper is organized as follows. The section II describes the Hamiltonian of the quantum system used in our analysis; for comparison of results, here we use  the same general form of the Hamiltonian as in \cite{fish}. The section III reviews the definition of thermodynamic limit for free energy and Fisher-Ruelle conditions on the non-random many-body potentials under which the free energy is extensive. To clarify our objectives from the onset,  this section also presents a statement of our results for the conditions in the case of disordered potentials.   The derivation of the conditions for both annealed as well as quenched disorder and for the finite and infinite limits of the system volume is described in section IV;  essentially being analogous to section III of \cite{fish}, the steps for infinite volume limit are mentioned only briefly (with some details given in {\it appendix A}).  In presence of the disorder, the spatial decay rate of the potential enters in the conditions through the  distribution of its random part and the results can vary based on the distribution parameters e.g. finite or infinite variances;  this is discussed in detail in sections V and VI. Our results clearly show a sensitivity of the thermodynamic limit to the nature of disorder, with latter often helping the LRIs to recover  their extensive behavior. Table I describes the parametric condition for five distribution types of the LRIs which leave the system extensive if fulfilled. Table II mentions the low temperature limit of the conditions on extensivity of LRIs.  An example illustrating our results is also discussed in {\it appendix B}. We conclude in section VIII with a brief discussion of the implications of our results.  

\section{Many body Hamiltonian}

Let $H({\bf p}_1,\ldots, {\bf p}_N; {\bf r}_1,\ldots, {\bf r}_N)$ be the Hamiltonian of  a quantum system  of volume $\Omega$ consisting of $N$ interacting ''particles'' (i.e sub-units) with their momenta and spatial coordinates as ${\bf p}_s, {\bf r}_s$, $s=1 \ldots ,N$. Assuming that the interacting part can be separated from the non-interacting one, $H$ can be written as 
\begin{eqnarray}
H = H_0 + U_N
\label{hg}
\end{eqnarray}
with $H_0({\bf p}_1,\ldots, {\bf p}_N; {\bf r}_1,\ldots, {\bf r}_N)$ as the total Hamiltonian of $N$ noninteracting "particles"
\begin{eqnarray}
H_0  =  \sum_{s=1}^{N}  H^{(s)}_0, 
\label{hg0} 
\end{eqnarray}
$ H^{(s)}_0 =H^{(s)}_0({\bf p}_s, {\bf r}_s)$ as the single-particle Hamiltonian of the particle labeled as ''$s$''
and $U_N \equiv U_N({\bf r}_1, {\bf r}_2,\ldots, {\bf r}_{N})$ as the total interaction among the particles.

In general, a many body potential among $N$ particles may consist of  the sum over contributions from $k$ body terms, with $1 \le k \le N$: 
\begin{eqnarray}
U_N =  \sum_{k=1}^N  U^{(k)} 
\label{vv} 
\end{eqnarray}
with $U^{(k)} $ as a $k$-body contribution 
\begin{eqnarray}
  U^{(k)} = \sum_{\{p\}} U^{(k,p)} ({\bf r}_{p1}, {\bf r}_{p2},\ldots, {\bf r}_{pk})
\label{vv1} 
\end{eqnarray}
with $\sum_p$ implying a summation over distinct $\left({\begin{array}{c} N  \\ k \end{array}}\right)$ combinations of $k$ particles chosen from the  set of  $N$ particles, with  subscript $p$ referring to one such combination and subscripts $p1, p2, \ldots, pk$ ranging from $1 \to N$.  Here we assume, as in \cite{fish}, that $U_N({\bf r}_1, {\bf r}_2,\ldots, {\bf r}_{N})$ is symmetric in $N$ variables ${\bf r_i}$, $i=1 \to N$. Note however, due to presence of disorder, $U_N$ is not translational invariant for our case.

For  application to real quantum systems, it is useful to assume $H$ to be a self-adjoint operator, thus implying it has real eigenvalues and a complete set of orthonormal eigenfunctions. As discussed in \cite{fish}, this assumption imposes constraints on the allowed boundary of the volume $\Omega$ and also requires the potential $U$ to be square-integrable. To proceed further, it is therefore necessary to define the domain confining the system.  Following the approach given in \cite{fish},  we consider a $d$-dimensional coordinate space, with position vectors ${\bf r}$, confined within a domain denoted by ${\mathcal D}$ and volume $\Omega=\Omega({\mathcal D})$. The domain is assumed to have a wall of thickness $h \ge 0$ so that the statement "${\bf r}$ is in ${\mathcal D}$" implies that the point ${\bf r}$ is at least at a distance $h$ from any boundary point of ${\mathcal D}$; this is equivalent to say that ${\bf r}$ is in a free volume $\Omega'$ where $\Omega' < \Omega$. 

For later reference, we also consider  two sub-domains  ${\mathcal D_1}, {\mathcal D_2}$ which may overlap but their free volumes are separated by the distance $R$ and lie within the free volume of domain ${\mathcal D}$. The sub-domains  ${\mathcal D_1}, {\mathcal D_2}$ are assumed to be of  volumes $\Omega_1, \Omega_2$ and contain $N_1, N_2$ particles respectively such that $\Omega=\Omega_1+\Omega_2$ and $N=N_1+N_2$. 
Consider $H_1({\bf p}_1,\ldots, {\bf p}_{N_1}; {\bf r}_1,\ldots, {\bf r}_{N_1})$ and $H_2({\bf p'}_1,\ldots, {\bf p'}_{N_1}; {\bf r}'_{1},\ldots, {\bf r}'_{N_2})$ as the Hamiltonians of these two parts which interact with each other with an interaction potential $\Phi$. Thus we have 
\begin{eqnarray}
 H = H_1 + H_2 + \Phi
\label{hh}
\end{eqnarray}
with  $H_1= \sum_{s=1}^{N_1} H_0^{(s)} + U_{N_1}$ and $H_2= \sum_{t=1}^{N_2} H_0^{(t)} + U_{N_2}$. Here $U_{N_1} =U_{N_1}({\bf r}_1, {\bf r}_2,\ldots, {\bf r}_{N_1}) $ corresponds to the interactions among the particles within domain ${\mathcal D}_1$ only. Similarly $U_{N_2} = U_{N_2} ({\bf r'}_{1}, {\bf r'}_{2},\ldots, {\bf r'}_{N_2})$ is related to the domain ${\mathcal D}_2$ only and $\Phi$ is the sum over those interactions of $U_N$ which are not contained in $U_{N_1}, U_{N_2}$ (i.e those consisting of particles from both volumes  $\Omega_1, \Omega_2$:  
\begin{eqnarray}
\Phi= \Phi ({\bf r}_1, {\bf r}_2,\ldots, {\bf r}_{N_1}, {\bf r'}_1, {\bf r'}_2,\ldots, {\bf r'}_{N_2})  
\label{nrwb}
\end{eqnarray}
Clearly the net potential energy $U_N = U_{N} ({\bf r}_1, {\bf r}_2,\ldots, {\bf r}_{N_1}, {\bf r'}_1, {\bf r'}_2,\ldots, {\bf r'}_{N_2})$ of the $N$ particles within 
domain ${\mathcal D}$ is the sum of the potential energies of the particles within domain ${\mathcal D}_1, {\mathcal D}_2$ and the interaction $\Phi$:  $U_N=U_{N_1} + U_{N_2} + \Phi$. 
Further note that
\begin{eqnarray}
\Phi= \sum_{k,l} \Phi^{(k,l)} =  \sum_{k,l,p,p'} \Phi^{(k,l,p,p')}.
\label{phi}
\end{eqnarray}
with superscripts $k, l$ implying  $k$  of them in domain ${\mathcal D}_1$ and $l$  of them in domain ${\mathcal D}_2$. Further $\sum_p$ and $\sum_{p'}$ refer to the summation over distinct  combinations of $k$ and $l$ particles, respectively, chosen from the  set of  $N_1$ and $N_2$ particles, respectively with  subscripts $p, p'$ referring to such combinations.  The number of $k+l$-body terms $\Phi^{(k,l)}$, with $k$  of them in domain ${\mathcal D}_1$ and $l$  of them in domain ${\mathcal D}_2$,  given as  
 \begin{eqnarray}
  M_{k+l}= \sum_{k,l}^{} {N_1 N_2 \over (l+1) (k-1)} \; 
\left(\begin{array}{c}  N_1-1 \\ l \end{array} \right)  \left(\begin{array}{c}  N_2-1 \\ k-1 \end{array} \right) 
\label{mk1}
\end{eqnarray}
which becomes very large in the thermodynamic limit (see appendix C of \cite{fish} for the derivation).

\section{Extensive nature of free energy: Conditions on potentials}

The  free energy $F$ of  a system, with Hamiltonian $H$ and at a temperature $T$, is defined as $F    =  -  {1\over \beta} \;   {\rm log} \; Z$ with $Z$ as  the canonical partition function $Z   =    {\rm Tr} \; {\rm e}^{-\beta H}$ and $\beta=(k T)^{-1}$. The thermodynamic limit of the free energy can be defined as follows \cite{fish}:
given a sequence of domains ${\mathcal D}_k, (k=0,1,2...)$ with volume $\Omega({\mathcal D}_k) \rightarrow \infty$ containing $N$ particles at fixed particle density $\rho$, the limiting free energy per particle, say $f =F/N$  becomes volume-independent:  
\begin{eqnarray}
\lim_{k \to \infty} f(\beta, \rho, \Omega_k) = f(\beta, \rho). 
\label{finf}
\end{eqnarray}
As discussed in \cite{fish}, the existence of the limit depends on two requirements as volume of the system increases (i)  a lower bound of the free energy per unit volume, say $f$, it should not diverge to $-\infty$, and (ii) an upper bound of the free energy per unit volume, that it does not diverge to $+\infty$. These bounds  on the free energy in turn manifest as  constraints on the many body potentials; here we state them first for clean potentials (derived  in \cite{fish})  and later on their generalization for disordered cases (derived later in this paper).

\subsection{Ruelle-Fisher Conditions on clean potentials}

As discussed in \cite{fish}, the bounds on free energy impose following constraints on the potentials: 

(a)   The lower bound on the potential, also referred as the stability condition, is given as  
\begin{eqnarray}
U_N({\bf r}_1, {\bf r}_2,\ldots, {\bf r}_{N}) \; \; \ge \; \; -  w_a \; N
\label{nrwa}
\end{eqnarray}
for all ${\bf r}_1, {\bf r}_2,\ldots, {\bf r}_{N}$ and for all $N$ with $w_a$ finite. The above relation is basically a statement about the stability of the system against its collapse due to attractive nature of the potential.  
More restrictive conditions ensuring thermodynamic limit can be also obtained for a class of stable potentials \cite{fish}.

(b) The mutual potential energy $\Phi(N_1, N_2)$ of the sets of $N_1$ and $N_2$  particles, separated from each other by a minimum distance $R$, satisfies the inequality, for some fixed $R_0$ and $w_b$, 
\begin{eqnarray}
\Phi({\bf r}_1, {\bf r}_2,\ldots, {\bf r}_{N_1}, {\bf r'}_1, {\bf r'}_2,\ldots, {\bf r'}_{N_2})  \; \; \le \; \;  {N_1\; N_2 \; w_b \over R^{d+\epsilon}} 
\label{nrwb}
\end{eqnarray}
if $|{\bf r}_i - {\bf r'}_j| \ge R \ge R_0$ for all $i=1, \ldots, N_1$ and $j=1, \ldots, N_2$  and ${(N_1+N_2)\over R^{d+\epsilon}} $ is sufficiently small with $\epsilon > 0$.  The above relation describes the stability of the system against the repulsive part of the many body interaction.  


\subsection{ Conditions on disordered potentials}

In presence of disorder, it is relevant to consider the thermodynamic limit of the disorder average (also referred as the ensemble average) of the free energy. The averaging (also referred as the ensemble average) however depends on the nature of the disorder i.e whether it is annealed or quenched: 
\begin{eqnarray}
\langle F \rangle   &=&   -  {1\over \beta} \;  \langle {\rm log} \; Z \; \rangle     \qquad {quenched} \\
&=& -  {1\over \beta}  \;  {\rm log} \langle \; Z \; \rangle  \qquad {annealed} 
\label{F}
\end{eqnarray}
with $\langle . \rangle$ implying a disorder average. (Here the annealed and quenched  disorder refer to system-dependence on random variables that do and don't evolve in time, respectively.  A quenched disorder average is therefore obtained by keeping the random variables fixed, while an annealed average is an average which is also carried out over all the possible values that the random variables can take).

Our objective in this paper is to derive the conditions on the disordered potentials for which 
$\langle F \rangle/ \Omega$ will have a well-defined thermodynamic limit. In this section, we state the conditions;  the details of their derivation are given  in section III and IV.

(a) The Hamiltonian $H$ for the domain  ${\mathcal D}$ represents a sufficiently well-behaved,  stable potential (system) so that  $\langle Z \rangle$ (quenched case) or $\langle \log Z \rangle$ (annealed case) exists. This in turn requires that on an average the minimum diagonal element, say  $U_{min}$, of potential $U$ in an arbitrary basis is bounded from below such that a finite $w_a$ (more accurately $w_a < \infty$) exists for all $N$ (equivalently volume $\Omega$ containing $N$ particles)  for which 
\begin{eqnarray}
-  {1\over \beta } \log\langle  {\rm e}^{- \beta \; U_{min} }\rangle  \; \;  & \ge \; \;   &
  - \; w_a \; N \qquad (annealed) \label{zq2} \\
\langle  U_{min} \rangle   \; \; & \ge \; \; &   - \; w_a \; N \qquad (quenched) 
 \label{zss2}
\end{eqnarray}

\vspace{0.2in}

(b) If one consider two domains say ${\mathcal D}_1$ and ${\mathcal D}_2$ separated from each other by a minimum distance $R$,  the interaction potential $\Phi$ of these domains must not depend too strongly on $N_1, N_2$ (alternatively their volume $\Omega_1, \Omega_2$) and must decay to zero with increasing $R$. Here $R$ is a length scale such that (i) $ |{\bf r}_i-{\bf r}'_j| \ge R$ for all particle-pairs $({\bf r}_i, {\bf r}'_j)$ with ${\bf r}_i$ in domain ${\mathcal D}_1$ and ${\bf r}'_j$ in ${\mathcal D}_2$, and, (ii) ${\Omega_1+\Omega_2 \over R^{d+\epsilon}}$ is sufficiently small  for a $d$-dimensional disordered system. The free energy  can be shown to be  bounded from above  if the largest diagonal, referred as  $\Phi_{max}$, of $\Phi$-matrix in an arbitrary basis (in which 
$H_1, H_2$ and $\Phi$ are statistically independent) satisfies following inequality, for all $N_1, N_2$, 
\begin{eqnarray}
-{1\over \beta} \; \log \langle  {\rm e}^{- \beta \; \Phi_{max}} \rangle & \; \; \le \; \; &   {  N_1 \; N_2  \; w_b \over  R^{d+ \epsilon}}    \hspace{1.1in}  (annealed) \label{zq3} \\
\langle  \Phi_{max}\rangle  \; \; & \le & \; \; {N_1 \; N_2 \; w_b \over  R^{d+\epsilon}}  \hspace{1.in} (quenched)
\label{zqq3}
\end{eqnarray}
where $w_b$ is finite.

As explained  later  in section V, VI, the conditions(\ref{zss2}, \ref{zq3}, \ref{zqq3}) can further be simplified, based on the tail behavior of the $\Phi_{max}$-distribution e.g. exponential or power-law (which governs the applicability of the central limit theorem) and the separability of its spatial dependence from  random degrees of freedom.

As clear from above,  in contrast to non-random case where the conditions for the  thermodynamic limits are on the potential itself, now only the distribution parameters are subjected to constraints

\section{Bounds on free energy in presence  of disorder} 

\subsection{ Lower bound on free energy}

Peirels theorem \cite{pe} states that  for a self adjoint operator $H$ 
\begin{eqnarray}
{\rm Tr} ( {\rm e}^{- \beta H}) =  \sum_{k}  \langle k | {\rm e}^{- \beta H} | k \rangle 
& \ge &  \sum_{k}  \; {\rm exp}\left[- \beta \langle k |  H | k \rangle \right]
\label{zp0}
\end{eqnarray} 
where  $|k \rangle$ is arbitrary basis.
Using the above, the partition function $Z(N,\Omega)  =  {\rm Tr} ( {\rm e}^{- \beta H})$ for the Hamiltonian $H=H_0+U$ can be written as

\begin{eqnarray}
Z(N,\Omega)   
& \ge & \sum_{k}  \; {\rm e}^{- \beta \;  (H_0)_{kk} }  \;  {\rm e}^{- \beta \;  U_{kk} }
\label{za2}
\end{eqnarray} 

Now let $U_{min}$ and $U_{max}$ be the minimum and maximum diagonals of the interaction potential $U$ in an arbitrary basis, then it can be shown that \cite{grif, fish}

\begin{eqnarray}
\left( {\rm Tr} \; {\rm e}^{-\beta H_0} \right)\; {\rm e}^{-{\beta} U_{min}}  \ge  Z(N,\Omega) \ge  \left({\rm Tr} \; {\rm e}^{-\beta H_0 } \right)\; {\rm e}^{-{\beta} U_{max}} 
\label{zp1}
\end{eqnarray} 

Using only the first inequality, one has 

\begin{eqnarray}
Z &\le &  Z_0   \; {\rm e}^{-{\beta} U_{min}} 
\label{zp2}
\end{eqnarray} 
where $Z_0 = {\rm Tr} \; {\rm e}^{-\beta H_{0}}$ is the partition function, with $H_0$ as the Hamiltonian for the system of  $N$ non-interacting particles confined within volume $\Omega $ with $\rho$ as the constant particle density: $N =\rho \; \Omega$.  

For clarity, let us assume that  $U_{min}$ corresponds to the $s^{th}$ diagonal of $U$:  $U_{min} \equiv U_{ss} = \langle s | U | s\rangle$.  For cases with $U$ given by eq.(\ref{vv}), one can write
\begin{eqnarray}
U_{ss} \equiv \sum_{k=1}^N \; U_{ss}^{(k)}   \hspace{1in}  U_{ss}^{(k)} = \sum_{p} \; U_{ss}^{(k,p)}
\label{emin}
\end{eqnarray} 
with $U_{ss}^{(k,p)} $ as the $s^{th}$ diagonal of the potential $U^{(k,p)}$.  Eq.(\ref{zp2}) can then be rewritten as 
\begin{eqnarray}
Z &\le &  Z_0   \; {\rm e}^{-{\beta} \; U_{ss}} 
\label{zp2x}
\end{eqnarray} 

The lack of interaction permits $Z_0$ to be  expressed in terms of the single particle partition functions $z_s$:  $Z_0 = (z_s)^N$ with $z_s={\rm e}^{-\beta H_0^{(s)}}$ with $H_0^{(s)}$ as the single particle Hamiltonian. To proceed further, we need to consider the  annealed or quenched disorder case separately.

\vspace{0.2in}

{\bf (i)  Annealed case:}  

\vspace{0.2in}

As the partition function $Z_0$ corresponding to non-interacting system is independent of the interaction potential, the ensemble average of both sides of eq.(\ref{zp2}) gives

\begin{eqnarray}
\langle Z \rangle 
&\le&  \langle Z_0 \rangle   \; \langle {\rm e}^{-{\beta} \; U_{min}} \rangle
\label{zp3}
\end{eqnarray} 

The above on substitution in eq.(\ref{F}) leads to 
\begin{eqnarray}
\langle  F(\Omega) \rangle  \ge   N \;  \langle F_s(\Omega) \rangle \; - {1\over \beta} \log\langle  {\rm e}^{-{\beta}  \; U_{min}} \rangle
\label{F1}
\end{eqnarray}
with $ F_s  = - {1\over \beta}  \log \; z_s  $ as the free energy of a single particle with $z_s$ as its partition function.  
If condition (\ref{zq2}) is now fulfilled, the  lower bound on $f$, the ensemble averaged free energy per particle for interacting case, becomes
\begin{eqnarray}
 f    \ge    \;   f_s - w_a
\label{F2}
\end{eqnarray}
where $f_s = \langle F_s \rangle$, is the ensemble-averaged free energy per particle for non-interacting case, or equivalently, the ensemble-averaged free energy for a single free particle. 
Clearly a finite lower limit of $f$ would then exist if $\omega_a$ remains finite in the infinite volume limit. Note $\omega_a$ can be temperature dependent but for the limit to exist at very low temperatures, $\omega_a$ should also be finite in $T \to 0$ limit. It is possible however that the approach to thermodynamic limit of a system varies with temperature.

\vspace{0.2in}

{\bf (ii) Quenched case} 

\vspace{0.2in}

First taking log of both sides of eq.(\ref{zp2}), followed by an ensemble average, gives 
\begin{eqnarray}
\langle  F(\Omega) \rangle  \ge   N\;  \langle F_s(\Omega_b) \rangle \; + \langle  U_{min} \rangle
\label{FF1}
\end{eqnarray}
Substitution of eq.(\ref{zss2})  in eq.(\ref{FF1}) now gives 
\begin{eqnarray}
 f    \ge    \;   f_s - w_a
\label{FF2}
\end{eqnarray}
Clearly a lower bound of $f$ exists if  the lower bound of ${\langle  U_{ss} \rangle}$ is given by eq.(\ref{zss2}), with a finite $w_a$ in the thermodynamic limit ($N, \Omega \to \infty$ with $\rho$ constant).

Note if ${\langle  U_{ss} \rangle \over N} \to 0$, the lower limit of the free energy of the interacting particles 
is then given by the non-interacting ones.  Clearly the lower limit of the free energy  exists for an arbitrary potential $U_N$ given by eq.(\ref{vv}), irrespective of the spatial range  of the many body terms $U^{(k)}$, as long as the minimum eigenvalues of the latter are   symmetrically distributed such that $\langle  U_{ss} \rangle=\langle  U_{ss}^{(k)} \rangle = 0$.

\subsection{Upper bound on free energy}

Following the approach of \cite{fish}, 
we now consider  a domain ${\mathcal D}$ of volume $\Omega$ containing $N$ particles divided into two sub-domains  ${\mathcal D_1}, {\mathcal D_2}$ 
which may overlap but their free volumes are separated by the distance $R$ and lie within the free volume of domain ${\mathcal D}$. The Hamiltonian 
in this case is given by eq.(\ref{hh}). 

Again applying Peirels's inequality to the partition function $Z(N,\Omega)  =  {\rm Tr} ( {\rm e}^{- \beta H})$ with $H$ given by eq.(\ref{hh}), we have in an arbitrary basis, say $| k \rangle$, 
\begin{eqnarray}
Z(N,\Omega)  
& \ge &  \sum_{k}  \; {\rm e}^{- \beta (H_1+H_2)_{kk}} \;\; {\rm e}^{- \beta \; \Phi_ {kk}}   \\
& \ge & {\rm e}^{- \beta \; \Phi_{max}}   \sum_{k}  \; {\rm e}^{- \beta (H_1+H_2)_{kk}} \;\; 
\label{zx2}
\end{eqnarray} 
where $\Phi_{max}$ is the largest diagonal of $\Phi$-matrix: $\Phi_{max} \ge  \Phi_ {kk}$ for all $k$. Henceforth subscript $\eta$ will be reserved for $\Phi_{max}$ i.e $\Phi_{\eta \eta} \equiv \Phi_{max}$. Note from eq.(\ref{phi})
\begin{eqnarray}
 \Phi_{\eta \eta} = \sum_{k} \sum_{l} \Phi^{(k,l)}_{\eta \eta},  \hspace{1in} \Phi^{(k,l)}_{\eta \eta}= \sum_{k,l,p,p'} \Phi^{(k,l, p, p')}_{\eta \eta}
\label{mdg}
\end{eqnarray}

\vspace{0.2in}

{\bf (iii) Annealed case:} 

\vspace{0.2in}

Assuming $H_1, H_2$ and $\Phi$ as statistically uncorrelated, 
the ensemble averaging then gives 
\begin{eqnarray}
\langle Z(N,\Omega)  \rangle
& \ge &  \sum_{k}  \; \langle {\rm e}^{- \beta (H_1+H_2)_{kk}} \rangle \;\langle  {\rm e}^{- \beta \; \Phi_{max}} \rangle
\label{zx4}
\end{eqnarray} 
To proceed further, let us write for simplification
$$\alpha =  {\Omega_1 \; \Omega_2 \;   w_b \over  R^{d+\epsilon}}.$$  
Now using eq.(\ref{zq3}), eq.(\ref{zx4}) can then be rewritten as 
\begin{eqnarray}
\langle Z \rangle   
& \ge & \langle Z_1 \rangle . \langle Z_2 \rangle . \;  {\rm e}^{-\beta \; \alpha}
\label{zff1}
\end{eqnarray}

Taking the logarithm of eq.(\ref{zff1}) and  using the definition for the ensemble averaged free energy per unit volume $f  = - {1\over \beta g}  \; \log \langle Z \rangle $   yields,  for both $w_b >0$ or $w_b <0$,
\begin{eqnarray}
\Omega \;  f    \; \; \le \; \;  \Omega_1 f_1  + \Omega_2 f_2  +  |\alpha | 
\label{zfa3}
\end{eqnarray}

By successive divisions of further domains ${\mathcal D}_3, {\mathcal D}_4$ from the domain ${\mathcal D}_1$ and iterating eq.(\ref{zff1}), we can obtain an inequality for an arbitrary 
subdivision of the original domain ${\mathcal D}$: 
\begin{eqnarray}
f(\rho,  \Omega)    
& \le & \sum_{m=1}^n  v_m \; f_m(\rho, \Omega_m)  + {1\over  \Omega}   \sum_{m=1}^{n-1} |\alpha_m |
\label{zf3}
\end{eqnarray}
where $v_m = {\Omega_m \over \Omega}$. Here again the free volumes of the $n$ sub-domains ${\mathcal D}_m$ are contained in the free volume of $\Omega$ but are separated from each other by at least the fixed distance $R$. Here the series in the last term comes because we gain additional terms $\alpha_n$ in successive stages:   
$\alpha_1=  {(\Omega- \Omega_2) \; \Omega_2  \; w_b \over  R^{d+\epsilon}}$, $\alpha_2=  {(\Omega- \Omega_2-\Omega_3) \; \Omega_3 \; w_b  \over R^{d+\epsilon} }$  and $\alpha_{n-1}=  {(\Omega- \sum_{j=2}^n\Omega_j) \; \Omega_n \;  w_b \over  R^{d+\epsilon}}$. As $\Omega \ge \sum_{j=2}^n \Omega_j$, one has 

\begin{eqnarray}
\sum_{m=1}^{n-1} |\alpha_m |  = 
\sum_{m=1}^{n-1} \; { (\Omega- \sum_{j=2}^{m+1} \Omega_j) \; \Omega_{m+1}}  {|w_b | \over  R^{d+\epsilon}} 
\hspace{0.2in}  \le \quad  { \Omega^2 \over  R^{d+ \epsilon}} \;  |w_b |. \nonumber \\
\label{zf4}
\end{eqnarray}
Substituting this in eq.(\ref{zf3}), we have, with  $\xi ={ \Omega \over R^{d+\epsilon}}$,  
\begin{eqnarray}
f(\rho, \Omega)    
& \le & \sum_{m=1}^n  v_m \; f_m(\rho, \Omega_m)  +  {|w_b |} \;  \xi
\label{zf5}
\end{eqnarray}

\vspace{0.2in}

{\bf (iv) Quenched case} 

\vspace{0.2in}
Proceeding from eq.(\ref{zx2}) by first taking $\log$ and then averaging, one can again arrive at 
eq.(\ref{zf5}) but now  $w_b$  is given  by the inequality (\ref{zqq3}). As clear, the condition is satisfied by  $w_b=0$, irrespective of the range of potentials, as long the disorder average of their off-diagonals is zero.

\vspace{0.2in}

\subsection{Thermodynamics limit and extensivity}

\vspace{0.2in}

 Eq.(\ref{zf5})  give the upper bound on the free energy per particle of the Hamiltonian $H$ for a disordered system of  volume $\Omega$ confined by a domain ${\mathcal D}$.
It is now relevant to consider the thermodynamic limit of the free energy i.e to analyze the form of its lower and upper bounds  in the limit $\Omega \rightarrow \infty, R \rightarrow \infty$ such that $\epsilon = {\Omega \over R^{d+\epsilon}} \to 0$.
Note eq.(\ref{zf5}) is essentially of the same form as eq.(5.5) of \cite{fish} (with following replacements  $N \to -f, \Omega \to {\mathcal D}, V \to \Omega$ where the symbols given on left of the $\to$ are those used in \cite{fish}).   Following the approach used in section 6 of \cite{fish}, 
the upper and lower bounds on free energy,  in large $k$ limit and for $\nu >d$, can be rewritten as  (details given in {\it appendix A})
\begin{eqnarray}
f(\rho, \Omega_k)  \; \;  \le \; \; f(\rho, \Omega_0)   + 
{|w_b | \;\xi_0 \; \varphi_2 \over (1-\varphi_2)}
\label{ze14}
\end{eqnarray}
with $\xi_0$ arbitrary, $\varphi_2 <1$ (see {\it appendix A}) and
\begin{eqnarray}
  f(\rho, \Omega_k)   \ge  \;  f(\rho, \Omega_0)    + w_a  
\label{ze16}
\end{eqnarray}

Here, as mentioned before,  $w_a, w_b$  must remain finite in the thermodynamics limit;  (note $w_a$ can  be a decreasing function of the volume). Further, analogous to case of non-random potentials too \cite{fish}, $w_a, w_b$ are temperature independent in  the quenched disorder case. However, for annealed case, the temperature-dependence of $w_a, w_b$ can not be ruled out.

As clear from eqs.(\ref{ze14},\ref{ze16}), an existence of finite $w_a, w_b$, satisfying conditions (\ref{zq2},\ref{zss2},\ref{zq3},\ref{zqq3}), in turn implies the existence of a free energy with upper and lower bounds in the thermodynamic limit.

\section{Role played by type of disorder: distribution with finite variance}

In presence of disorder,  each of the $k$-body  contributions $U^{(k,p)}$ (eq.(\ref{vv1})) and $\Phi^{(k,p)}$ are randomized, with their matrix elements behaving like random variables  if the basis to represent them  is chosen appropriately e.g. the eigenfunction basis of the Hamiltonian in absence of disorder. With $U_{ss}$ and $ \Phi_{\eta \eta}$ given by eqs.(\ref{emin},\ref{mdg}) respectively, both  of them behave as random variables too. Based on the nature of randomness and mutual dependence of various terms contributing to them, the conditions can be rewritten in terms of the distribution parameters which gives better insight about their applicability. 

For later reference, an important point worth emphasizing here is following. As the question regarding an existence of upper bound of free energy is concerned with  repulsive core of a potential at large particle-distances, the matrix elements of $\Phi$ in any physically meaningful basis are expected to be positive. Further, as $\Phi$ describes the interaction between two domains at a spatial distance $R$, this results in a $R$-dependence of $\Phi_{\eta \eta}$ and thereby its distribution parameters. Intuitively the mean and variance of the distribution for $\Phi_{\eta \eta}$ are expected to have the same spatial-dependence as its typical value.


 For cases in which $U^{(k,p)}$ for $k=2 \to N$ are mutually independent,  both $U^{(k)}_{ss} =  \sum_{p} U_{ss}^{(k,p)}$ as well as 
 $\Phi^{(k,l)}_{\eta \eta} = \sum_{p,p'} \Phi_{\eta \eta}^{k,l,p,p'}$ are  summations over  many independent random variables. With  number of terms contributing to $U^{(k)}_{ss}$ and $\Phi^{(k,l)}_{\eta \eta}$ becoming very large in the thermodynamic limit, the standard central limit theorem (CLT)  predicts their  distribution to approach Gaussian limit. The latter helps as the averages  in eqs.(\ref{zq2},\ref{zq3}) can then be simplified by following identity for a Gaussian random variable, say $y$ with mean $u$ and variance $\sigma^2$
\begin{eqnarray}
\langle  {\rm e}^{- \beta \; y} \rangle  ={\rm e}^{(1/2) \beta^2 \sigma^2 - \beta u }
 \label{clt1}
\end{eqnarray}
Consider that the diagonal element $U_{ss}^{(k,p)}$ of $U^{(k,p)} ({\bf r}_{p1}, \ldots, {\bf r}_{pk})$ are distributed  with mean $u^{(k,p)}$ and variance $\sigma^{2(k,p)}$. Following CLT, the  mean $u_k$ and variance $\nu_k^2$ for the Gaussian distributed  $U^{(k)}_{ss}$ can be expressed as
\begin{eqnarray}
u_k =  \sum_p u^{(kp)},  \hspace{0.5in} \sigma_k ^2 = \sum_p \sigma^{2(kp)}
\label{unu}
\end{eqnarray}
Similarly assuming that $\Phi^{(k,l,p,p')}_{\eta \eta}$  is  distributed  with mean $\mu^{(klpp')}$ and variance $\nu^{2(klpp')}$, the  mean $\mu_{kl}$ and variance $\nu_{kl}^2$ for the Gaussian distributed  $\Phi^{(k,l)}_{\eta \eta}$ can be expressed as
\begin{eqnarray}
\mu_{kl} =  \sum_{p,p'} \mu^{(klpp')},  \hspace{0.5in} \nu_{kl}^2 = \sum_{p,p'} \nu^{2(klpp')}
\label{unu1}
\end{eqnarray}

 Further assuming that many body interactions $U^{(k)}$ for different $k$ are mutually independent,  the latter would also be applicable for their diagonals $U^{(k)}_{ss}$. Applying the same reasoning, maximum diagonals $\Phi^{(k,l)}_{\eta \eta}$ for different $k,l$ can also be assumed independent. Following eq.(\ref{emin}) and eq.(\ref{mdg}), this implies
\begin{eqnarray}
\langle U_{min} \rangle \equiv \langle U_{ss} \rangle &=& \sum_k \langle U_{ss}^{(k)} \rangle, \label{e2} \\
\langle \Phi_{max} \rangle  \equiv \langle \Phi_{\eta \eta} \rangle &=& \sum_{k,l} \langle {\Phi}_{\eta \eta}^{(kl)} \rangle. \label{p2}
\end{eqnarray}
and
 \begin{eqnarray}
\langle  {\rm e}^{-{\beta} \; U_{min}} \rangle \equiv \langle  {\rm e}^{-{\beta} \; U_{ss}} \rangle  = \prod_{k}  \langle  {\rm e}^{-{\beta} \;  U^{(k)}_{ss}} \rangle  \label{e1} \\
\langle  {\rm e}^{-{\beta} \; \Phi_{max}} \rangle \equiv \langle  {\rm e}^{-{\beta} \; \Phi_{\eta \eta}} \rangle  = \prod_{k,l} \langle  {\rm e}^{-{\beta} \;  \Phi^{(k,l)}_{\eta \eta}} \rangle  \label{p1}
\end{eqnarray}

A point  worth emphasizing here is as follows:  from eq.(\ref{emin}) and eq.(\ref{mdg}), $U_{ss}$ as well as $\Phi_{\eta \eta}$ can directly be written as the sum over  $U^{(kp)}_{ss}$ and $\Phi^{(klpp')}_{\eta \eta}$ respectively  which suggests one to apply CLT directly  to $U_{ss}$ as well as $\Phi_{\eta \eta}$ . But note 
$U^{(kp)}$ for different $k$ values refer to the interactions corresponding to different numbers of particles and in general need not be identically variables; (a similar argument can be extended to  $\Phi^{(klpp')}_{\eta \eta}$ too).  The CLT in its standard form is however applicable to a sum over iid variables. Although many generalized variations of CLT applicable to non-iid variables are available in scientific literature, they are often applicable under specific restrictions on the nature of randomness of the variables. For generic considerations, it is therefore more appropriate to apply CLT to $U_{ss}^{(k)}$ as well as $\Phi^{(k,l)} _{\eta \eta}$.

To proceed further, we consider annealed and quenched cases separately.

\subsection{Annealed case}

\vspace{0.2in}

\noindent{{\bf Lower Bound}  
Applying the relation (\ref{clt1}) for $y \to U^{(k)}_{ss}$  gives 
$ \langle  {\rm e}^{-{\beta} \;  U^{(k)}_{ss}} \rangle  = {\rm e}^{(\beta^2/2) \nu_{kl}^2 - \beta \mu_{kl}}$.
 The latter  on substitution in eqs.(\ref{e1}) leads to
\begin{eqnarray}
\langle  {\rm e}^{-{\beta} \; U_{ss}} \rangle  & =&  { \rm exp}\left[\sum_{k} \left({\beta^2 \sigma_k^2\over 2}  - \beta u_k \right)\right] \label{unv} 
\end{eqnarray}
with $u_k, \sigma_k$ defined in eq.(\ref{unu}).
With help of the above, eq.(\ref{zq2})  can then be rewritten as 
\begin{eqnarray}
{1\over N} \; \sum_{k=1}^N \left(u_k - {\beta \over 2 } \sigma_k^2 \right) \;  \ge \;  - \; {w_a }
\label{con1a}
\end{eqnarray}
where $w_a$ is finite but arbitrary otherwise. 
 Further defining  $u = {1\over N} \; \sum_{k=1}^N \; u_k$ and $\sigma^2 = {1\over N} \; \sum_{k=1}^N \; \sigma_k^2$,  eq.(\ref{con1a}) can be simplified as 
 \begin{eqnarray}
u - {\beta \over 2 } \sigma^2 \;  \ge \;  - \; {w_a }
\label{com1}
\end{eqnarray}
 
Note $u$ and $\sigma^2$ correspond to an average of mean values and  variances, respectively, of all many body contributions to the potential $U$. Consequently, for the cases with Gaussian decay with finite mean and variance, $u$ and $\sigma^2$ are expected to be finite and a finite $w_a$ can always be found.  Following eq.(\ref{ze16}), this in turn implies that, at finite temperature, a lower limit of average free energy can always be defined for Gaussian distributed many body potentials. 
But at low temperature near $T \to 0$,  condition(\ref{com1}) can not be satisfied unless  $\sigma^2$ also varies with temperature (e.g. $\sigma^2 \sim {1\over \beta}$); note however in the latter case the condition reduces to almost same form as in the case of non-random potentials.

\vspace{0.2in}

\noindent{\bf Upper Bound} 
Applying  the relation (\ref{clt1}) to $ \Phi^{(k,l)}_{\eta \eta}$ gives $\langle  {\rm e}^{-{\beta} \Phi^{(k,l)}_{\eta \eta}} \rangle  ={\rm e}^{(\beta^2/2) \nu_{kl}^2 - \beta \mu_{kl}}$ with  $\mu_{kl}, \nu_{kl}$ defined in eq.(\ref{unu1}).
Substitution of the latter in eqs.(\ref{p1}) gives
\begin{eqnarray}
\langle  {\rm e}^{-{\beta} \Phi_{\eta \eta}} \rangle  &=& {\rm exp}\left[\sum_{k,l} \left({\beta^2 \nu_{kl}^2\over 2}  - \beta \mu_{kl} \right)\right] \label{pnv}
\end{eqnarray}
Using eq.(\ref{pnv}) in eq.(\ref{zq3}) 
then leads to 
\begin{eqnarray}
\sum_{k,l=1}^{N_1 N_2} \left({\mu_{kl}} - {\beta \over 2} {\nu_{kl}^2} \right) \le {N_1 N_2 w_b \over R^{d+\epsilon}}.
\label{cg1}
\end{eqnarray}

Further defining $\mu = {1\over N_1 N_2} \sum_{k,l=1}^{N_1 N_2} {\mu_{kl}}$ and  $\nu = {1\over N_1 N_2}\sum_{k,l=1}^{N_1, N_2} {\nu_{k,l}}$, the above inequality can be rewritten as
\begin{eqnarray}
{\mu} - {\beta \over 2} {\nu^2}  \le { w_b \over R^{d+\epsilon}}.
\label{cg2}
\end{eqnarray}

For cases  with  $2 \mu \le {\beta \sigma^2}$,  the condition (\ref{cg2}) is satisfied for $w_b =0$, (the left side of eq.(\ref{cg2}) being negative-definite as $\nu^2 \ge 0$). Consequently, following eq.(\ref{ze14}),  an upper limit of free energy exists, for finite temperatures, for any $d$-dimensional disordered many body potential of arbitrary spatial  decay if $2 \mu \le {\beta \sigma^2}$. Further, even if 
$\mu \sim {1\over R^{\gamma}} >0$ with $\gamma$ arbitrary, eq.(\ref{cg2}) is satisfied for very low temperatures ($\beta \to \infty$) irrespective of $R$-dependence of $\nu$.  
In opposite case of $2 \mu > {\beta \sigma^2}$, a finite $w_b$ can again be defined if $\mu \sim {1\over R^{\gamma}}$ with $\gamma > d$. Clearly in this case, the condition for existence of upper limit is same as in the case of clean potentials.   

As clear from the above, a  competition between mean and variance,  latter dominating the former with  help of low temperature, fulfills the condition for upper limit for potentials with arbitrary spatial decay.

\subsection{Quenched case}

 To determine the upper and lower bounds in this case,  only a knowledge of mean values   $\langle U_{ss} \rangle$ and $\langle \phi_{\eta \eta} \rangle$ is needed. As discussed above, $U_{ss} $  behaves as a product of Gaussian variables $U_{ss}^{(k)} $, with its  mean given by eq.(\ref{e2}); the condition (\ref{zss2}) can then be rewritten as 
\begin{eqnarray}
 u  \ge - w_a.
\label{FF+2}
\end{eqnarray}
where $u$ is same as defined above eq.(\ref{com1}).
Clearly, $u$ being finite, the above condition can be fulfilled for an arbitrary potential $U$ irrespective of its spatial range. Similarly $\phi_{\eta \eta} $ behaves as a product of Gaussian variables $\phi_{\eta \eta}^{(kl)} $, with its  mean given by eq.(\ref{p2}); 
the condition (\ref{zqq3}) for upper limit can then be written as
\begin{eqnarray}
{\mu}  \le { w_b \over R^{d+\epsilon}}.
\label{cg3}
\end{eqnarray}
with $\mu$ again same as defined above eq.(\ref{cg2}). Clearly, if $\mu >0$ ($\Phi$ being repulsive potential), a finite $w_b$ exists if $\mu \sim {1\over R^{\gamma}}$ with $\gamma > d$ which is analogous to the corresponding  condition for clean potentials.   
Clearly, contrary to annealed Gaussian potentials, the quenched Gaussian disorder does not help the extensive nature of long range interactions.  

To clarify the above results, an example  for both annealed as well as quenched cases with Gaussian disorder is discussed  in {\it appendix} B. 

\subsection{Clean limits}
For variance-limits $\sigma_k^2, \nu_{kl}^2 \to 0$, the Gaussian distribution of variables $U_{ss}^{(k)}$  and $\phi^{(k,l)}_{\eta \eta}$ reduce to Dirac-delta functions $\delta(U_{ss}^{(k)} - u_k)$ and 
$\delta( \phi^{(k,l)}_{\eta \eta}- \nu_{kl})$(peaked at corresponding mean values $u_k$ and $\mu_{kl}$ and zero elsewhere). Clearly  the  $u=\sum_k u_k$ and $ \mu=\sum_{k,l} \mu_{kl}$ in these limits are equivalent to $U_{ss}$  and  $\phi_{\eta \eta}$ and  the conditions in eq.(\ref{com1}), eq.(\ref{cg2}), eq.(\ref{FF+2}), eq.(\ref{cg3}) reduce to corresponding limits  for clean quantum systems discussed in \cite{fish}.

\section{Role played by type of disorder: distributions with Power-law tails}

 Many physical variables e.g. many body potentials  often reveal a  stable distribution with asymptotic power law decay which corresponds to infinite variance. A stable distribution in general is described by four parameters, say $a,b,c,\delta$ referred as the stability, skewness, scale  and location parameters of the distribution, respectively, and can be defined as \cite{stab}
 \begin{eqnarray}
f(x; a, b, c,\delta) &=& {1\over \pi} \; {\bf Re} \int_0^{\infty} {\rm e}^{i t(x-\delta)} {\rm e}^{-(c t)^{a} \; (1-i b \phi)} \; {\rm d}t,  \label{fs1} 
\label{f1}
\end{eqnarray}
with $\phi= \tan(\pi a/2)$ for $a \not=1$, $\phi=-(2/\pi) \log|t|$ for $a=1$; (note another expression for $\phi$ is also used sometimes \cite{stab}:  $\phi= ((ct|^{1-a} -1) \tan(\pi a/2)$ for $a \not=1$, $\phi=-(2/\pi) \log|ct|$ for $a=1$). Here the parameters are confined within following ranges:
\begin{eqnarray}
a \in (0,2], \quad b \in [-1,1], \quad c \in (0, \infty), \quad \delta \in (-\infty, \infty)
\label{abcd}
\end{eqnarray}
with support of the distribution depending on $a, b$: 
\begin{eqnarray}
&& x \in (-\infty, \infty) \; \;  {\rm if} \; \; b \not=\pm 1, \nonumber \\  
&& x \in \left[\delta-c \; \tan(\pi a/2), \infty \right) \; \; {\rm if} \; \;  a <1, b=1 \nonumber \\  
&& x \in \left(-\infty, \delta + c \; \tan(\pi a/2) \right] \; \; {\rm if} \; \; a <1, b=-1. 
\label{sup}
\end{eqnarray}

A relevant point for comparison with non-random cases is that, in the limit $a \to 0$ or $c \to 0$, the variable $x$ described by stable distribution approaches its non-random limit: $f(x;a,b,c,\mu) \to \delta(x-\mu)$.

As examples and also for later reference, we mention here three important stable distributions, namely Levy ($a=1/2, b=1$ and $ x \in \left[\delta, \infty \right)$), Pareto ($ x \in (-\infty, \infty)$) and Cauchy ($a=1, b=0$ and $ x \in (-\infty, \infty)$), with their probability densities given as follows (with subscripts $L,P,C$ on $f$ referring to Levy, Pareto or Cauchy distribution, respectively) \cite{stab}:
 \begin{eqnarray}
{\rm {\bf Levy}} \hspace{0.22in} f_L(x; c, \delta) \equiv  f(x; 1/2,1,c,\delta) &=& \sqrt{c \over 2 \pi} (x- \delta)^{-3/2} {\rm e}^{-{c\over 2(x-\delta)}}, \label{levy} \\
{\rm {\bf Pareto}} \hspace{0.2in} f_P(x; a, c) \equiv f(x; a,b,c,\delta) &=& {a c^a \over x^{a+1}}  \hspace{0.3in}   (x \ge c),  \hspace{0.3in} = 0  \hspace{0.3in}   (x < c)   \label{Pareto} \\
{\rm {\bf Cauchy}} \hspace{0.2in} f_C(x; c, \delta) \equiv f(x;1,0,c,\delta) &=& {1\over \pi c}  \left[{c^2 \over c^2 + (x-\delta)^2 } \right]
 \label{Cauchy}
\end{eqnarray} 

\vspace{0.2in}

{\bf Evaluation of Averages:}
As mentioned in previous section, the standard  central limit theorem is applicable for a sum of  independent and identically distributed (iid) random variables with finite variances.   For cases where the random variable is described by a non-degenerate stable distribution with power law tails, a generalized central limit theorem can be invoked \cite{stab}: consider random variables $x_n$, $n=1 \to N$ distributed with probability density $f(x_n; a,b_n,c_n,\delta_n)$. The generalized CLT (GCLT) predicts that the sum 
\begin{eqnarray}
y = \sum_{n=1}^N x_n
\label{y}
\end{eqnarray}
 will tend to a stable distribution $f(y; a,b,c,\delta)$ as the number of random variables grows where
 \begin{eqnarray}
c^a &=& \sum_{n=1}^N  c_n^a, \hspace{0.3in} b =  c^{-a} \;  \sum_{n=1}^N  b_n \; c_n^a  \nonumber \\
\delta &=&\sum_n \delta_n +\tan(\pi a/2) \left(b c - \sum_{n=1}^N b_n c_n \right)   \hspace{0.5in} a \not=1  \nonumber \\
&=& \sum_n \delta_n + {2\over \pi} \left(b \; c \log c - \sum_{n=1}^N b_n c_n \log c_n \right)   \hspace{0.3in} a =1
\label{f2}
\end{eqnarray}
For the case in which $x_n$ are independent and identically distributed say with density $f(x_n,a,b_0,c_0,\delta_0)$, $y$ approaches the distribution described by $f(y; a,b,c,\delta)$ with $b=b_0$, $c^a=N c_0^a$, $\delta =  N \delta_0 
+ N b_0 c_0 \tan(\pi a/2) \left(N^{(1-a)/a}-1 \right)$ for $a \not=1$ and  $\delta_N =  N \delta 
+ {2 \over \pi} b_0 c_0 N \log N$  for $a =1$.

The calculation of the averages is easier for cases with symmetric  stable distribution $f(y,a,0,c,\delta)$  and $\beta >0$ (later referred as sym-st). The averages  can however be defined only in a restricted region    $\delta < u \le y \le \infty$;  eq.(\ref{f1}) 
gives (using $b=0$) 
 \begin{eqnarray}
 \langle  {\rm e}^{-{\beta} y} \rangle_{S,res}  &=&   {1\over \pi}  \; \sum_{n=1}^{\infty} {(-1)^{n+1} (c \beta)^{an} \over n!} \;  \; \sin\left({n a  \pi\over 2}\right) \; \Gamma(an+1) \;  \Gamma \left(-a n, \beta(u-\delta)\right)  \; {\rm e}^{-\beta \delta}  \label{eya} \\
 \langle  y \rangle_{S,res}  &=&   {1\over \pi} \;  \sum_{n=1}^{\infty}  {c^{an}  \over n!} \; {(a n u -\delta) \; \Gamma{(an-1)}\over  (u -\delta)^{an}}  \; \cos\left({n a\pi\over 2}\right)  
\label{yaa}
\end{eqnarray}
with notation $\langle . \rangle_{S,res}$ implying an ensemble average over the  restricted region in which such averages can be defined.

It is more instructive to consider the cases with special values of $a,b,c, \delta$.
 As mentioned above, with $x_n$ given by the distribution $f_L(x_n; c_n, \delta_n)$, $f_P(x_n; a, c_n)$ or $f_C(x_n, c_n, \delta_n)$, the GCLT  predicts $y$ to be distributed as $f_L(y;c,\delta)$, $f_P(y,a,c)$  or $f_C(y; c, \delta)$, respectively, with $c, \delta$ given by eq.(\ref{f2}); ($c=\left(\sum_n \sqrt{c_n} \right)^2, \delta= \sum_n \delta_n + \left( \sum_n \sqrt{c_n} -\sum_n c_n \right)$ for Levy case, $c^a = \sum_n c_n^a$ for Pareto and $c =\sum_n c_n, \delta= \sum_n \delta_n$ for Cauchy cases).
 Using eqs.(\ref{levy}, \ref{Pareto}, \ref{Cauchy})  for the distribution of $y$, the averages can then be given as
\begin{eqnarray}
 \langle  {\rm e}^{-{\beta} y} \rangle_L  &=&  {\rm e}^{- \beta \delta - \sqrt{2 \beta c} } \label{yal} \\
 \langle  {\rm e}^{-{\beta} y} \rangle_P   &=& 
  {a c^a \beta^a}  \; \Gamma{(-a, \beta c)}   \label{yap} \\
   \langle  {\rm e}^{-{\beta} y} \rangle_{C,res}   &=&   {1\over 2} \left[{\rm e}^{i \beta c} \Gamma{(i \beta c)} +
   {\rm e}^{- i \beta c} \Gamma{(- i \beta c)} \right]\label{yac}
\end{eqnarray}
 with  $\langle \rangle_L, \langle \rangle_P, \langle \rangle_C$ referring to an averaging over Levy, Pareto or Cauchy distributed $y$, respectively.  Note here eq.(\ref{yac}) is valid only for partial averaging i.e for $\delta \le y \le \infty$ instead of entire support of Cauchy distribution (i.e $-\infty \le y \le \infty$); this is equivalent to considering only a part of the ensemble of Cauchy distributed $y$.
  
 Similarly 
\begin{eqnarray}
 \langle {y} \rangle_{L,res}  &=& {1\over 2\sqrt{\pi}} \left[c \; \Gamma{(-1/2, t/2)} + 2 \; \delta \; \Gamma{(1/2, t/2)} \right]  \label{yal1} \\
 \langle  { y} \rangle_P   &=&   {a \; c \over a-1}   \hspace{0.1in}  (a >1), 
 \hspace{0.1in} = \infty \hspace{0.1in}  (a \le 1)     \label{yap1} \\
   \langle { y} \rangle_C   &=& \delta \label{yac1}
\end{eqnarray}
where the relation in eq.(\ref{yal1}) is valid for the cases with a  finite upper limit of $y$ (i.e only for partial averaging if $\delta \le y \le (c+t \delta)/t$, with $t >0$, instead of entire support).

\subsection{Annealed case}

\vspace{0.2in}

As examples of annealed disorder with stable distribution, here we consider four cases mentioned above. With eqs.(\ref{e1},\ref{p1}) still applicable for the averages, the lower and upper bounds $w_a, w_b$ can then be obtained by using eqs.(\ref{eya},\ref{yal},\ref{yap},\ref{yac}) as follows.

\vspace{0.2in}

\noindent{\bf Lower Bound:}
Assuming that $U_{ss}^{(k,p)}$ is described by a non-degenerate stable distribution $f(U_{ss}^{(k,p)}, a_k, b_{kp}, c_{kp}, \delta_{kp})$, the above, along with eq.(\ref{emin}), then implies that $U^{(k)}_{ss}$ approaches a stable distribution $f(U_{ss}^{(k)}, a_k, b_{k}, c_{k}, \delta_{k})$ with  
its parameters given by  eq.(\ref{f2}) (with replacements $b \to b_k, c \to c_k, \delta \to \delta_k$ in the left side of the equation and $b_n \to b_{kp}, c_n \to c_{kp}, \delta_n \to \delta_{kp}$ in the right side). 
Using eqs.(\ref{yal}, \ref{yap}, \ref{yac}) for 
$y \to U^{(k)}_{ss}$ , followed by eq.(\ref{e1}) gives $\langle {\rm e}^{-\beta U_{min}}\rangle$. The latter on substitution in eq.(\ref{zq2}) then leads to the condition 
\begin{eqnarray}
{\overline X}_L \equiv  {1\over N} \sum_{k=1}^N   X_k  \; \; & \ge & \; \;   - {w_a } 
\label{zxl0}
\end{eqnarray}
with  $X_k =X(a_k, b_k, c_k, \delta_k)$  where 
\begin{eqnarray}
X(a,b,c,\delta) &=& {\delta}  + \sqrt{2 c \over \beta}   
 \hspace{3.42in} {\rm {\bf {Levy}}} \label{zqwl1} \\
 &=&  {-1 \over  \beta} \log \left[a \;  \left({\beta c}\right)^a  \;  \Gamma{\left(-a, {\beta c} \right)} \right] \hspace{2.12in} {\rm {\bf {Pareto}}}  \label{zqwp1}\\  
&=& \delta +  {\log 2\over \beta} -{1\over \beta} \rm log \left[{\rm e}^{i \beta c} \Gamma{\left({i \beta c}\right)} + {\rm e}^{- {i \beta c} } \Gamma{\left(- {i \beta c} \right)} \right]  
 \hspace{0.9in}  {\rm {\bf {Cauchy}} }  \label{zqwc1}\\
&=& \delta -  {1\over  \beta} \log\left( 
  {1\over \pi}  \; \sum_{n=1}^{\infty} {(-1)^{n+1} (\beta c)^{n a } \over n!  }  \;  \; \sin\left({n a \pi \over 2}\right) \; \Gamma(n a+1) \;  \Gamma \left(-n a, {\beta (u-\delta)} \right)  \right)  \nonumber  \\ 
 & & \hspace{4.12in}  {\rm {\bf {SymSt}}}   \label{zqws1}
\end{eqnarray} 
with $u$ defined above eq.(\ref{eya}) and
\begin{eqnarray}
&&c_k=\left(\sum_p \sqrt{c_{kp}} \right)^2, \hspace{0.3in} \delta_k= \sum_p \delta_{kp} + \left( \sum_p \sqrt{c_{kp}} -\sum_{p} c_{kp} \right) \hspace{0.8in} {\rm \bf(Levy)}, \nonumber \\, 
&&\delta_k = \sum_p \delta_{kp} \hspace{4.in}  \;\; {\rm \bf(Pareto)}, \nonumber \\ 
&&c_k =\sum_{p} c_{kp}, \hspace{0.8in} \delta_k= \sum_{p} \delta_{kp} \hspace{2.4in} {\rm \bf(Cauchy)}, \nonumber \\
&&c_k =\left(\sum_{p} c_{kp}^{a_k}\right)^{1\over a_k}, \hspace{0.3in} \delta_k= \sum_{p} \delta_{kp} \; {\rm with} \; 0 < a_{k} <2,  \hspace{1.3in}  {\rm \bf(SymSt)} 
\label{pk}
\end{eqnarray}

 Note, as mentioned above, eq.(\ref{zqwc1}) and eq.(\ref{zqws1}) are applicable only for restricted support (for $\delta_k \le U_{ss}^{(k)} \le \infty$,   and $\delta_k < u  \le U_{ss}^{(k)} \le \infty$, respectively).

As  the left side of eq.(\ref{zxl0}) is a combination of many parameters,  they may conspire together, for some cases,  to give rise to a  finite $w_a$. For example, table I illustrates the parametric combinations for which ${\overline X}_L \ge 0$, thus satisfying the condition(\ref{zxl0}), with $w_a=0$, even for arbitrary spatial dependence of distribution parameters. As another example, consider the low temperature limit ($\beta \to \infty$) of eq.(\ref{zxl0}). With definitions
\begin{eqnarray}
\overline{\delta}_L \equiv {1\over N} \sum_{k=1}^N  \delta_k, \qquad {\overline c}_L \equiv \left({1\over N} \sum_{k=1}^N {c_k}^{a_k} \right)^{1\over a_k}.
\label{del}
\end{eqnarray}
  eq.(\ref{zxl0}) can now be approximated as 
\begin{eqnarray}
{\overline \delta}_L   \ge   - {w_a  } \;  {\rm \bf(Levy)},\;\;  {\overline c}_L   \ge   - {w_a  } \; {\rm \bf(Pareto)}, \;\; {\overline \delta}_L +{\pi \over 2}{\overline c}_L   \ge   - {w_a  } \; {\rm \bf(Cauchy)},\; \; u_L   \ge   - {w_a  } \; {\rm \bf(symst)}
\nonumber
\end{eqnarray}
 Clearly, in low temperature limit,  $w_a$ exists for Levy, Cauchy  and sym-stable cases if $\overline{\delta}_L$ is  finite (as ${\overline c}_L >0$ and $u > \delta_L$, see eq.(\ref{abcd})). For Pareto case however the above limit can always be satisfied  e.g for $w_a=0$. This becomes more clear by an example with iid variables, discussed in {\it appendix} B.

\vspace{0.2in}

\noindent{\bf Upper Bound:} 
For $\Phi^{(k,l,p,p')}_{\eta \eta}$ distributed as 
$f(\Phi^{(k,l,p,p')}_{\eta \eta}; a_{kl}, b_{klpp'}, c_{klpp'}, \delta_{klpp'})$, here again GCLT implies that $\Phi^{(k,l)}_{\eta \eta}$  given by eq.(\ref{mdg}) approaches the distribution $f(\Phi^{(k,l)}_{\eta \eta}; a_{kl}, b_{kl}, c_{kl}, \delta_{kl})$ with its parameters given by eq.(\ref{f2}) (following replacements $b \to b_{kl}, c \to c_{kl}, \delta \to \delta_{kl}$ in the left side of the equation and $b_n \to b_{klpp'}, c_n \to c_{klpp'}, \delta_n \to \delta_{klpp'}$ on its right side). 
For $f$ corresponding to Levy, Pareto or Cauchy  distributions, 
the upper limit $w_b$ can then be obtained as follows: using 
eqs.(\ref{eya}, \ref{yal}, \ref{yap}, \ref{yac}) for $y = \phi^{(kl)}_{\eta \eta}$, followed by its substitution in eq.(\ref{p1}), gives $\langle {\rm e}^{-\beta \phi_{max}} \rangle=\langle {\rm e}^{-\beta \phi_{\eta \eta}} \rangle$. The latter on substitution in  eq.(\ref{zq3})  gives
\begin{eqnarray}
\overline{X}_U \equiv {1\over N_1 N_2} \sum_{k,l=1}^{N_1 N_2} X_{kl}   \; \; & \le & \; \;   {w_b \over R^{d+\epsilon}}.   \label{zxu0}
\end{eqnarray} 
where  $X_{kl} = X(a_{kl}, b_{kl}, c_{kl}, \delta_{kl})$ with $X$  given by eqs.(\ref{zqwl1}, \ref{zqwp1}, \ref{zqwc1}, \ref{zqws1}).  
Here
\begin{eqnarray}
&&c_{kl}=\left(\sum_{p,p'} \sqrt{c_{klpp'}} \right)^2, \; \; \delta_{kl}= \sum_{p,p'} \delta_{klpp'} + \left( \sum_{p,p'} \sqrt{c_{klpp'}} -\sum_{p,p'} c_{klpp'} \right) \; \; {\rm \bf(Levy)}, \nonumber\\
&&\delta_{kl} = \sum_{p,p'} \delta_{klpp'} \hspace{3.5in}  \;\; {\rm \bf(Pareto)}, \nonumber \\ 
&&c_{kl} =\sum_{p,p'} c_{klpp'}, \delta_{kl}= \sum_{,p',p} \delta_{klpp'} \; \hspace{2.5in} {\rm \bf(Cauchy)}, \nonumber \\
&& c_{kl} =\left(\sum_{p,p'} c_{klpp'}^{a_{kl}}\right)^{1\over a_{kl}} \; {\rm with} \; 0 < a_{kl} <2,  \hspace{1.9in}  {\rm \bf(SymSt)} 
\label{pkl}
\end{eqnarray} 
Note, as mentioned in previous section, the distribution parameters of $\Phi^{(k,l)}_{\eta \eta}$ can be $R$-dependent, $\Phi^{(k,l)}$ being the interaction between two domains at a minimum distance $R$. Here again the results for Cauchy and sym-stable distributions are applicable for restricted support only.

Once again, due to appearance of multiple parameters on its left side, the condition in  eq.(\ref{zxu0}) has the possibility of fulfillment irrespective of the spatial dependence of the distribution parameters. For example  one such case is the parametric conditions for which ${\overline X}_U \le 0$ (with details given in Table I).  Another useful example is  the large $\beta$-limit of eq.(\ref{zxu0}).  Using definitions
 \begin{eqnarray}
 {\overline\delta}_U \equiv {1\over N_1, N_2} \sum_{k,l=1}^N  \delta_{kl}, \qquad {\overline c}_U \equiv \left({1\over N_1 N_2} \sum_{k,l} c_{kl}^{a_k}\right)^{1 \over ^{a_k}}. 
 \label{del1}
 \end{eqnarray}
 eq.(\ref{zxu0}) can now be approximated as 
\begin{eqnarray}
{\overline \delta}_U \le {w_b \over R^{d+\epsilon}} \; {\rm \bf(Levy)},  \;\;
{\overline c}_U    \le   {w_b \over R^{d+\epsilon}  } \; {\rm \bf(Pareto)}, 
\;\;
{\overline \delta}_U + {\pi \over 2} {\overline c}_U    \le   {w_b \over R^{d+\epsilon}  } \; {\rm \bf(Cauchy)}, 
\;\; u \le {w_b \over R^{d+\epsilon}} \; {\rm \bf(SymSt)},
\nonumber 
\end{eqnarray}
  Further recalling that $u > {\overline \delta}_U$ and $ {\overline \delta}_U$ can be negative (see eq.(\ref{abcd}) and the text above eq.(\ref{eya})), the above condition can be satisfied by Levy and sym-stable distributions for the cases with ${\overline \delta}_U < 0$ even if $|{\overline \delta}_U| \sim {1\over R^{\gamma}}$  for arbitrary $\gamma$. 
But as ${\overline c}_U \ge 0$ (see eq.(\ref{abcd})), Pareto distribution fulfills the above condition only if ${\overline c}_U =0$ or ${\overline c}_U \sim {1\over R^{\gamma}}$ with $\gamma >d$. In Cauchy case, however, an  additional presence of ${\overline \delta}_U$ in the bound may help to overcome the positive definite contribution from ${\overline c}_U$ e.g if both ${\overline \delta}_U \sim {-\alpha_0\over R^{\gamma}}, {\overline c}_U \sim {\alpha_1\over R^{\gamma}}$ with $\alpha_0 > \alpha_1 >0$ even if $\gamma < d$. Clearly the Pareto type disorder  does not help LRIs to attain the thermodynamic limit but the disorder of Levy, Cauchy  or symmetric stable types can.

\subsection{ Quenched case}

As in  the annealed case discussed above, here again $U_{ss}^{(k)}$ and $\phi_{\eta \eta}^{(kl)}$  approaches the same stable distributions as that of  $U_{ss}^{(kp)}$ and $\phi_{\eta \eta}^{(klpp')}$, respectively, with relation between their parameters given by eq.(\ref{pk}) and eq.(\ref{pkl}).  But a determination of  $w_a, w_b$ now requires a knowledge of mean values $\langle U_{ss}^{(k)}\rangle$ and $\langle \phi_{\eta \eta}^{(kl)} \rangle$ only 
which can be obtained by eq.(\ref{e2}, \ref{p2}). As examples, here again we give the results for quenched disorder with Levy, Pareto, Cauchy or symmetric-stable distributions.

\vspace{0.2in}

{\bf Upper Bound:}  Using eq.(\ref{p2}) along with eqs.(\ref{yaa}, \ref{yal1}, \ref{yap1}, \ref{yac1}) for 
$y = \phi^{(kl)}_{\eta \eta}$, followed by its substitution in eq.(\ref{zq2}), the condition (\ref{zqq3}) now becomes 

 \begin{eqnarray}
\overline{Y}_U \equiv {1\over N_1  N_2} \sum_{k,l} Y_{kl}    \; \; & \le & \; \;  { w_b \over R^{d+\epsilon}},   \label{ydu} 
\end{eqnarray}
with $Y_{kl} \equiv Y(a_{kl}, b_{kl}, c_{kl},\delta_{kl})$ where
\begin{eqnarray}
Y(\alpha, \xi, \gamma, \eta) 
&=& {1\over 2 \sqrt{\pi} } \left[ \gamma \; \Gamma{\left(-{1\over 2}, {t\over 2} \right)} + 2 \eta \; \Gamma{\left({1\over 2}, {t\over 2} \right)}\right]  \approx  {\gamma-\eta \over 2 \sqrt{\pi} } \left[(1+  {t^2 (\gamma -2 \eta)\over \sqrt{\pi} (\gamma-\eta) } \right]  \hspace{0.0in} {\rm {\bf {Levy}}}  \label{ydl1} \\
&= & {\alpha \; \gamma \over 1-\alpha}      \hspace{0.3in}  (\alpha >1),      \hspace{0.3in}   = \infty  \hspace{0.3in}  (\alpha \le 1) \hspace{.7in} {\rm {\bf {Pareto}}}  \label{ydp1} \\
&= &  {\eta }    \hspace{3.42in} {\rm {\bf {Cauchy}} }\label{ydc1} \\
&= & \sum_{n=1}^{\infty} { (\gamma)^{n \alpha}  \over n!} \; {(n u \alpha -\eta) \; \Gamma{(n \alpha-1)}\over  (u -\eta)^{n \alpha}}  \; \cos\left({\pi n \alpha\over 2}\right)  \hspace{0.40in} {\rm {\bf {SymSt}}}
\label{yds1}
\end{eqnarray}
with eq.(\ref{ydl1}) and eq.(\ref{yds1}) applicable for restricted support only (i.e for $\delta_{kl} \le \phi_{\eta \eta}^{(kl)} \le (c_{kl}/t + \delta_{kl})$ and $\delta_{kl} \le u \le \phi_{\eta \eta}^{(kl)} \le \infty$). 
Here again, $a_{kl}, b_{kl}, c_{kl},\delta_{kl}$  are given by eq.(\ref{pkl}) and can in general be a function of spatial distance $R$ between the domains. Clearly,  in case of a potential $\Phi$ with arbitrary spatial range $R^{-\gamma}$, at least one way to approach the upper limit is if $\overline{Y}_U \le 0$. The parametric conditions  in  which the latter can be achieved are illustrated in Table I.

 \vspace{0.2in}

{\bf Lower Bound:} Again using eqs.(\ref{yal1}, \ref{yap1}, \ref{yac1}, \ref{yaa}) for 
$y \equiv U^{(k)}_{ss}$ with replacements $c \to c_k, \delta \to\delta_k $, followed by 
eq.(\ref{e2}) and  its substitution in eq.(\ref{zss2}), then gives the condition
\begin{eqnarray}
\overline{Y}_L \equiv {1\over N} \sum_{k} Y_{k}  \; \; & \ge & \; \; -w_a.    \label{ydl}
\end{eqnarray}
Here $Y_k =  Y(a_k, b_k, c_k, \delta_k)$ for each of the four cases is given by eqs.(\ref{ydl1}, \ref{ydp1}, \ref{ydc1}, \ref{yds1}) but with $a_k, b_k, c_k, \delta_k$ now given by eq.(\ref{pk}). Clearly for the parametric conditions leading to  a finite 
$\tilde{Y}_L$, a finite value of $w_a$ can always be found. As an example, Table 1 gives, for the four cases, the parametric 
conditions which lead to $\overline{Y}_L \ge 0$ and thereby satisfy the condition eq.(\ref{ydl})  for $w_a=0$.

The tables I and II summarize our results for the five distribution types mentioned above. We further elucidate our ideas by an example discussed in {\it appendix} B. 

\subsection{Clean limits}
It is  worth recalling  that, the limits $a \to 0$ or $c \to 0$ correspond to the clean (non-random) limit of  the distribution $f(x;a,b,c,\delta)$ of the variable $x$ (as $f$ is peaked around $x=\delta$ and zero elsewhere). A substitution of $a_k \to 0$ or $c_k \to 0$ in eqs.(\ref{zqwl1}, \ref{zqwp1}, \ref{zqwc1}, \ref{zqws1}) then leads to the clean limits of  eq.(\ref{zxl0}) and eq.(\ref{ydl}): ${\overline \delta}_L \ge w_a$. Similarly
substituting $a_k \to 0$ or $c_{kl} \to 0$ gives the  clean limits of eq.(\ref{zxu0}) and eq.(\ref{ydu}): ${\overline \delta}_U \le {w_b \over R^{d+\epsilon}}$. As expected,  the  clean limits are same for both annealed and quenched cases and, with replacements $\delta_k \to U^{k}$, $\delta_{kl} \to \phi^{(k,l)}$, coincide with results given in \cite{fish} for clean systems,   (also given by eqs.(\ref{nrwa}, \ref{nrwb}) along with eqs.(\ref{vv}, \ref{vv1}, \ref{phi})).

\section{Conclusion}

To understand the role of disorder, we analyzed the extensive limits for a number of prototypical 
disordered many-body potentials. Our results reveal that disorder  often helps quantum systems to attain the thermodynamic limit  by relaxing the conditions on the spatial range of potentials. While for non-random cases the need for extensivity  imposes constraints directly on each realization of the potential, in contrast the conditions in presence of disorder are only on the the average/ typical average of the disordered potential and its moments. This indicates that even though not all realizations of the potential may individually satisfy the extensivity requirement, its fulfillment on an average across the disordered ensemble is sufficient. This is useful because the conditions on the distribution parameters of complicated potentials can be more easily fulfilled as the volume increases. Under certain parametric condition, this helps to reduce the lower limit on the spatial range of  "extensive" interactions.
In this context, our analysis  reveals the crucial role played by the nature of disorder i.e annealed vs quenched in attaining thermodynamic limit. The conditions in case of an annealed disorder turn out to be temperature-sensitive,  a fingerprint of the underlying dynamics which equilibrates itself  with changing temperature. For low enough temperatures and based on the type of distribution of the potential (more specifically, its diagonal matrix element in the physically relevant basis), the distribution parameters can conspire together to fulfill the condition necessary for the existence of upper bound of free energy (a statement on the repulsive nature of the potential) even if the  potential is spatially long-ranged (spatial decay of the potential is slower than the physical dimensions of the system); Tables I and II describe the parametric conditions for the existence of extensive limit for five prototypical  distributions. Although we have confined here to quantum potentials and canonical ensemble, our results can  be generalized to classical systems as well as to grand canonical ensembles; (as mentioned before, similar results have been known in context of classical long-range lattice models \cite{ks, fz, z, eh, e}).

As suggested by previous studies of complex systems,  the role of non-homogenized, local interactions is akin to that of disorder, at least  in context of  the statistical properties. Thus we expect our results to be applicable also for a clean system with varying range of interactions across a single sample.   It seems the complexity, irrespective of its origin, helps to locally block the interactions at far-parts, effectively making them shorter range so that they can achieve thermodynamic limit and stability.

\acknowledgements

I am indebted to Professor Anthony Leggett for motivating me to pursue this analysis and for many discussions on the subject. I am also grateful to Professor Michael Berry for helpful conversations.

\appendix

\section{Derivation of eq.(\ref{ze14}) and eq.(\ref{ze16})}

\vspace{0.2in}

In section IV.B, we derived  the upper bound on the free energy per particle of the Hamiltonian $H$ for a disordered system of  volume $\Omega$ confined by a domain ${\mathcal D}$. As obvious, the upper bound is the sum of the free energies of the sub-volumes contained in $\Omega$ but all of them separated from each other by a minimum distance $R$; (here $R$ is the length scale such that $ |R_s-R_t| \ge R$ for all pairs of $(s,t)$ particle-pairs with $s$ in domain ${\mathcal D}_1$ and $t$ in domain ${\mathcal D}_2$).  As discussed in \cite{fish}, this minimum distance is basically to take ito account the thickness of the wall of each of the volumes which however approaches zero in infinite volume limit.

Our next step is to consider the thermodynamic limit of the free energy i.e to analyze the form of its lower and upper bounds  in the limit $\Omega \rightarrow \infty, R \rightarrow \infty$ such that $\xi = {\Omega \over R^{d+\epsilon}} \to 0$.
Note eq.(\ref{zf5}) is essentially of the same form as eq.(5.5) of \cite{fish} (with following replacements  $g \to -f, \Omega \to {\mathcal D}, V \to \Omega$ where the symbols given on left of the $\to$ are those used in \cite{fish}).

Following the approach used in section 6 of \cite{fish}, we consider a sequence of cubic domains ${\mathcal D}_k, (k=0,1,2,\ldots$) of edge $a_k$ with volumes $\Omega_k$  and the wall-thickness $h_k$.  Now assuming that the edge of the cube at $(k+1)^{th}$ step of the sequence is twice that of at $k^{th}$ step, one has $a_k= 2^k a_0$ and the nominal volume $\Omega_k = a_k^d = 2^{kd} \; a_0$. Both $\Omega_k$ and $h_k$ are assumed  to increase to infinity in a way such that $\xi_k$ and the fraction of the volumes excluded by the walls tend to zero;  this can be done by  assuming the wall-thickness to be just a small fraction of the edge of the cube: $h_k = b_k \; a_k$  with fractional thickness $b_k= \varphi_1^k \alpha_0$ with $1/2 < \varphi_1 < 1$ so that $\lim_{k \to \infty} b_k \to 0$ while $\lim_{k \to \infty} h_k \to \infty$.     As described in \cite{fish}, a cubic domain  ${\mathcal D}_{k+1}$ at $(k+1)^{th}$ sequence-step consists of $2^d$ cubic domains ${\mathcal D}_{k}$, with their free volumes lying within the free volume of ${\mathcal D}_{k+1}$ but separated from each other by a distance 
\begin{eqnarray}
R_{k+1} = 2[h_k - (h_{k+1}-h_k)] = 4(1-\varphi_1) (2 \varphi_1)^k \; h_0 .   
\label{rrx}   
\end{eqnarray}
As clear $R_{k+1} > R_0$  if $h_0$ is chosen large enough. Now by defining $\varphi_2 = 2^{(d-\nu)/2} \; \varphi_1^{- \nu} < 1$, the repulsion parameter can now be 
rewritten as
\begin{eqnarray}
\xi_{k+1} = {\Omega_{k+1} \over R_{k+1}^{\nu}} = \xi_0 \; \varphi_2^{k+1} .
\label{rpx}      
\end{eqnarray}
Thus $\xi_k \to 0$ as $k \to \infty$. Note the condition $\varphi_2 <1$ can be fulfilled by choosing the $\varphi_1= 2^{(d-\nu)/2\nu}$ with $\nu > d$ which also satisfies the assumption made above i.e $\varphi_1 <1$.

Let $f({\mathcal D}_k)=f_k$ be the free energy density at stage $k$. Then application of the basic inequality (\ref{zf5})  with two sets of $4$ cubes (each of volume $\Omega_k$)  leads to
\begin{eqnarray}
f_{k+1}(\rho) - |\omega_b| \; \xi_{k+1}   
& \le &  {1\over 2} \; f_{k,1}(\rho)  + {1\over 2} \;  f_{k,2}(\rho)   
\label{zf6x}
\end{eqnarray}
But as the cubes at step $k$ are all identical, the above equation can be rewritten as 
\begin{eqnarray}
f_{k+1}(\rho) -  |\omega_b| \; \xi_{k+1}   & \le &  f_{k}(\rho)    
\label{zf7x}
\end{eqnarray}

Subtraction of $t_k \equiv  { |\omega_b| } \sum_{n=0}^{k} \xi_n $ from both the sides gives 
 \begin{eqnarray}
  f_{k+1}(\rho) - t_{k+1}    & \le &  f_k(\rho) - t_k
\label{zf8x}
\end{eqnarray}

Now using eq.(\ref{rpx}), we have 
\begin{eqnarray}
t_k =   |w_b| \; \xi_0 \sum_{n=0}^k \varphi_2^n =
{ |w_b | \; \xi_0 \; (1-\varphi_2^{k+1}) \over (1-\varphi_2)}
\label{tk}
\end{eqnarray}
 which implies $\lim_{k \to \infty} t_k \rightarrow { |w_b| \; \xi_0\over (1-\varphi_2)}$. Thus if  we define $q_k \equiv f_k - t_k$, then eq.(\ref{zf8x}) gives $q_k$ as a monotonically decreasing  sequence but bounded from below through eq.(\ref{F2}), that is 
\begin{eqnarray}
q_{k+1} \le q_k
\end{eqnarray} 
As $q_k$ is a decreasing function with respect to $k$, its limit is bounded from above by any $q_M$ with $M < k$: $q_{\infty} \le q_k \le q_{k-1} \le \ldots  \le  q_{2}  \le q_1 \le q_0 $.  

Using now $q_k \le q_M$ for all $M \le  k$ gives the upper bound on the free energy 
 \begin{eqnarray}
f(\rho, \Omega_k) \le f(\rho, \Omega_M)  -  (t_M-t_k)
\label{zf12x}
\end{eqnarray}
for all $M < k$. But as $$ t_k-t_M  =  | w_b | \; \xi_0 \; \sum_{n=M+1}^k  \varphi_2^n  ={ |w_b| \; \xi_0 \; \varphi_2^{M+1} (1-\varphi_2^{k-M}) \over (1-\varphi_2)},$$ 
taking $M=0$, we have 
 \begin{eqnarray}
f(\rho, \Omega_k) \le f(\rho, \Omega_0)  +  { |w_b| \; \xi_0 \; \varphi_2  (1-\varphi_2^{k}) \over (1-\varphi_2)}
\label{zf13x}
\end{eqnarray}
which can be rewritten as 
 \begin{eqnarray}
f(\rho, \Omega_k) - 
{|w_b |  \;\xi_0 \; \varphi_2 \over (1-\varphi_2)}  \le f(\rho, \Omega_0)  - { |w_b| \; \xi_0 \; \varphi_2^{k+1} \over (1-\varphi_2)} 
\label{zf14x}
\end{eqnarray}

But now using $q_{\infty} = f_{\infty} - t_{\infty}$, with $f_{\infty}$ having a lower bound given by eq.(\ref{F2}), 
along with $q_k \ge q_{\infty}$, we can write the lower bound on $q_k$:
 \begin{eqnarray}
q_k \ge  \;  f(\rho, \Omega_0)   -  {1\over \beta \Omega} \log\langle  {\rm e}^{-{\beta} U_{min}} \rangle  -  {|w_b| \; \xi_0 \over (1-\varphi_2)}
\label{zf11x}
\end{eqnarray}
Using now $q_k=f_k -t_k$ on the lhs of eq.(\ref{zf11x}) and rearranging gives
\begin{eqnarray}
  f(\rho, \Omega_k)  +   {1\over \beta \Omega_k} \log\langle  {\rm e}^{-{\beta} U_{min}} \rangle \ge  \;  f(\rho, \Omega_0)    -  {|w_b| \; \xi_0 \; \varphi_2^{k+1} \over (1-\varphi_2)}
\label{zf16x}
\end{eqnarray}
with help of eq.(\ref{zq2}), the above inequality can be rewritten as 
\begin{eqnarray}
  f(\rho, \Omega_k)   \ge  \;  f(\rho, \Omega_0)    -  {|w_b| \; \xi_0 \; \varphi_2^{k+1} \over (1-\varphi_2)} + w_a
\label{zf16y}
\end{eqnarray}

Now as  $\varphi_2 < 1$ if $\nu >d$, this implies $\lim_{k \to \infty} \; (\varphi_2)^k  \rightarrow 0$.  In large $k$ limit and for $\nu >d$, therefore, eq.(\ref{zf14x}) and eq.(\ref{zf16y}) can be rewritten as 
\begin{eqnarray}
f(\rho, \Omega_k)  \; \;  \le \; \; f(\rho, \Omega_0)   + 
{|w_b | \;\xi_0 \; \varphi_2 \over (1-\varphi_2)}
\label{ze14x}
\end{eqnarray}
and
\begin{eqnarray}
  f(\rho, \Omega_k)   \ge  \;  f(\rho, \Omega_0)    + w_a  
\label{ze16x}
\end{eqnarray}
Here, as mentioned before,  $w_a, w_b$  must remain finite in the thermodynamics limit;  (note $w_a$ can  be a decreasing function of volume). 
Further, analogous to case of non-random potentials too \cite{fish}, $w_a, w_b$ are temperature independent in  the quenched disorder case. However, for annealed case, the temperature-dependence of $w_a, w_b$ can not be ruled out.


\section{ Example: Two-body interaction with a random and a non-random component}

Consider a system with its $g$ particles interacting via a pair-wise coupling of random single particle fields represented by an operator $\Lambda$. The  Hamiltonian of the system can be given by eq.(\ref{hg}) with the potential $U$ as   
\begin{eqnarray}
U =  \;  \sum_{s,t=1 \atop s \not=t}^{N}  
\;  \;  {\Lambda^{(st)}  \over   | \; {\bf r}_s-{\bf r}_t \; |^{p} }.   
\label{vg}
\end{eqnarray}
Choosing an arbitrary  $N$-dimensional fixed basis $|k \rangle$, $k=1\to N$, the matrix elements of $U$ can be given as 
\begin{eqnarray}
U_{kl} =  \sum_{s,t=1 \atop s \not=t}^{N}  
\;  \;  { \Lambda^{(st)}_{kl} \over   | \; {\bf r}_s-{\bf r}_t \; |^{p} }\; \;  \; \;   
\label{vgekl}
\end{eqnarray}

Following the definition  of   $\Phi$ given by eq.(\ref{phi}), its maximum diagonal element, required to determine $w_b$,  can be given as  
\begin{eqnarray}
\Phi_{max} = \Phi_{\eta \eta} = \sum_{s=1}^{N_1} \; \sum_{t=1}^{N_2} \;  \;  {  \Lambda^{(st)}_{\eta \eta}  \over   | \; {\bf r}_s-{\bf r'}_t \; |^{p} }
\label{vge12}
\end{eqnarray}
Let us now define $\Lambda_0$ as follows:
$\Lambda_0 =  \sum_{s=1}^{N_1} \; \sum_{t=1}^{N_2} \;  | \Lambda^{(st)}_{\eta \eta}  |$. The latter along with eq.(\ref{vge12}) gives 
\begin{eqnarray}
\Phi_{max}   <  {\Lambda_0 \over R^p}
\label{vgg}
\end{eqnarray}
with $R$ as the minimum distance between the free volumes of the domains $\Omega_1, \Omega_2$ i.e  $R < | {\bf r_s}-{\bf r_t}|$ for all $(s,t)$-pairs (as defined in section II). The above leads to

\begin{eqnarray}
-{1\over \beta} \; \log \langle  {\rm e}^{- {\beta \Phi_{max}}} \rangle &  \; \; \le \; \; &
 -{1\over \beta} \; \log \langle  {\rm e}^{- {\beta \Lambda_0 \over R^p}} \rangle
 \label{aaq3} \\
 \langle  \Phi_{max} \rangle &  \; \; \le \; \; &  {\langle \Lambda_0 \rangle \over R^p}
 \label{abq3} 
\end{eqnarray}
Here, as  $\Lambda_0$ is a sum over a large number of iid positive random variables $| \Lambda^{(st)}_{\eta \eta}  |$, each say with mean $\lambda$ and variance  $\eta$, one can invoke CLT to calculate the averages on the left side.

To find $w_a$ for this case, we again need a prior information about minimum eigenvalue of $U$. Let $\lambda^{(st)}_{min}$ be the minimum eigenvalue of the randomized pair-interaction $\Lambda^{(st)}$. As assumed above, the latter are independent for different pairs which implies $V_{min}$ as the sum over large number of independent random variables:  
\begin{eqnarray}
U_{min}  >  \sum_{s,t=1}^N  {\lambda_{min}^{(st)} \over |{\bf r_s} - {\bf r_t}|^{p}}   >  -  {\lambda_{min} \over L^p} \; .
\label{vmin}
\end{eqnarray}
where $\lambda_{min} = \sum_{s,t=1}^N  |\lambda_{min}^{(st)}|$ and   $L$ be  the largest possible distance between particles in a given volume $\Omega$: $|{\bf r_s} - {\bf r_t}| \le  L$.
The above gives
\begin{eqnarray}
-  {1\over \beta } \log\langle  {\rm e}^{- \beta \; U_{min} }\rangle  \; \;  & \ge \; \;   & 
-  {1\over \beta } \log\langle  {{\rm e}^{\beta \lambda_{min} \over L^p} }\rangle
\label{aaq4} \\
\langle  U_{min} \rangle  \; \;  & \ge \; \;   & - { \langle \lambda_{min}\rangle \over L^p}
\label{abq4}
\end{eqnarray}

Further evaluation of inequalities (\ref{aaq3}, \ref{aaq4}) depends on the type of randomness of the variables $\Lambda_0$ and $\lambda_{min}$. Here we again consider the distributions with finite and infinite variances separately.
 
\vspace{0.1in}

{\bf Annealed distribution with finite variance:} Assuming $\Lambda^{(st)}_{\eta \eta}$ as iid random variables with mean $\mu_0$ and finite variance $\nu_0$ for all $\{s,t\}$ pairs, the CLT predicts $\Lambda_0$ to approach a  Gaussian distribution with mean $\mu = N_1 N_2 \mu_0$ and variance $\nu^2 = N_1 N_2 \nu^2_0$; eq.(\ref{clt1}) then implies $\langle  {\rm e}^{- {\beta \Lambda_0 \over R^p}} \rangle  ={\rm e}^{-\beta ( { \mu \over R^{p}} - {\beta \nu^2  \over R^{2p}})}$. The latter along with eq.(\ref{aaq3}) gives the upper bound
\begin{eqnarray}
-{1\over \beta} \; \log \langle  {\rm e}^{- {\beta \Phi_{max}}} \rangle &  \; \; \le \; \; &
 N_1 N_2  \; \left( { \mu_0 \over R^p} - {\beta \nu^2_0 \over R^{2p}} \right) 
\end{eqnarray}

The condition (\ref{zqq3}) for the upper limit on free energy can then be fulfilled if a finite $w_b$ can be defined such that
\begin{eqnarray}
  \; \left( { \mu_0 \over R^p} - {\beta \nu_0^2 \over R^{2p}} \right) &  \; \; \le \; \; &
  { w_b \over  R^{d+ \epsilon} }    \qquad  (annealed) \label{ap3} 
\end{eqnarray}
For the temperatures $T \to 0$, when the 2nd term on the left side of the above equation  dominates (note both $\nu, \mu$ and $R >0$), the condition can be fulfilled with $w_a=0$  irrespective of power $p$ of the interaction. For finite $T$ too, a finite $w_a$ exists even for $p < d$  if $\mu \to 0$ . Clearly, near zero temperatures or symmetrically distributed disordered potential (\ref{vg}), an upper limit of the free energy exists irrespective of the spatial dependence of the potntial (i.e even for $p < d$ with $d$ as the physical dimension of the system).
This is in contrast to clean systems where the upper limit of free energy exists, in general, for short range interactions i.e those spatially decaying faster than volume of the system.

For $\Lambda^{(st)}$ for various $s,t$-pairs as iid random interactions, their minimum eigenvalues $\lambda_{min}^{(st)}$ are iid random variables, say with mean   $u_0$ and variance $\sigma_0^2$. 
Following the central limit theorem,  the distribution of $\lambda_{min}$ in the large volume limit can again be given by the Gaussian, with mean $N u_0$ and variance $N \sigma_0^2$. 
Using the above, eq.(\ref{aaq4})  can then be rewritten as 
\begin{eqnarray}
-  {1\over \beta } \log\langle  {\rm e}^{- \beta \; U_{min} }\rangle  \; \;  & \ge \; \;   &  - N \left({u_0 \over L^p} + {\beta \over 2 } {\sigma_0^2\over L^{2p}} \right) 
\label{con1}
\end{eqnarray}
A comparison with eq.(\ref{zq2}) now indicates that $w_a$ can be defined in terms of $u_0$ and $\sigma^2_0$: $w_a = {u_0\over L^p} + {\beta \over 2 }{ \sigma_0^2 \over L^{2p}} $. With $L \approx {\mathcal S} \; \Omega^{1/d}$, with ${\mathcal S}$ as a shape-dependent positive constant, $w_a \to 0$ for  finite temperature $T$. For $T \to 0$ however, existence of a finite $w_a$ depends on the competition of limits $\Omega \to \infty$ and $\beta \to \infty$;  for  $T L^p \to 0$, it is possible again to define a finite $w_a$ ($w_a \to 0$).

\vspace{0.1in} 

{\bf Annealed, power law distributions}: again assuming $\Lambda^{(st)}_{\eta \eta}$ as iid random variables distributed with probability density  $f(\Lambda^{(st)}_{\eta \eta}; a_0, b_0, c_0,\delta_0)$ with $f$ given by Levy, Pareto or Cauchy distribution, the GCLT  predicts $\Lambda_0$ to be distributed as $f(\Lambda_0; a, b, c,\delta)$, respectively; 
here $a=a_0=1/2, b=b_0=1, c=c_0 (N_1 N_2)^2$  and $\delta=N_1 N_2 \delta_0$ for Levy,  $a=a_0, c=N_1 N_2 c_0$ for Pareto, $a=a_0=1, b=b_0=0, c = N_1 N_2 c_0$ and $\delta=N_1 N_2 \delta_0$ for Cauchy.  
Substituting eqs(\ref{yal1},\ref{yap1},\ref{yac1}) with $y=\Lambda_0$ and $\beta \to {\beta \over R^p}$ in eq.(\ref{aaq3}), the condition for the upper limit can  be given as follows

\begin{eqnarray}
{X\over N_1 N_2}   \; \; & \le &   {w_b \over R^{d+\epsilon}}. \label{zx} 
 \end{eqnarray}
with $X \equiv X(a,b,c,\delta)$ where $X$ is defined as 
\begin{eqnarray}
X(a,b,c,\delta) &=& {\delta \over R^p}  + \sqrt{2 c \over \beta R^p}   
 \hspace{3.02in} {\rm {\bf {Levy}}} \label{zqwl2} \\
 &=&  {-1 \over  \beta} \log \left[a \;  \left({\beta c\over R^p}\right)^a  \;  \Gamma{\left(-a, {\beta c \over R^p} \right)} \right] \hspace{1.5in} {\rm {\bf {Pareto}}}  \label{zqwp2}\\  
&=&  \delta + {\log 2\over \beta} -{1\over \beta} \rm log \left[{\rm e}^{i \beta c \over R^p} \Gamma{\left({i \beta c\over R^p}\right)} + {\rm e}^{- {i \beta c \over R^p} } \Gamma{\left(- {i \beta c \over R^p} \right)} \right]  
 \hspace{0.4in}  {\rm {\bf {Cauchy}}}   \label{zqwc2}
\end{eqnarray}

For Levy case, the condition (\ref{zx}) can be simplified as 
 ${\delta_0 \over R^p}  + \sqrt{2 c_0 \over \beta R^p}   \le {w_b \over R^{d+\epsilon}}$. Clearly a finite $w_b$ in large $R$  limit  can be found for arbitrary $p$ if only $\delta_0 < 0$ and $\beta$ is large. For Pareto case,    
the condition can be approximated as ${c_0 \over R^p}   \le {w_b \over R^{d+\epsilon}}$ (neglecting the contribution from logarithmic terms); as $c_0 > 0$, a finite $w_b$ now exists only for $p \ge d+\epsilon$. 
Similarly, for Cauchy case, a finite $w_b$ for arbitary $p$ exist if $\delta_0$ is sufficiently negative. 
Thus the condition (\ref{zqq3}) for the upper bound on free energy can be fulfilled for a random  potential with long range spatial decay (i.e $p < d$) if it is Levy or Cauchy distributed but not in the case of Pareto distribution.

Again assuming $\lambda^{(st)}_{min}$  as iid distributed with probability density  $f(\lambda^{(st)}_{min};  \tilde{a_0},\tilde{b_0}, \tilde{c_0},\tilde{\delta_0})$ for all $\{s,t\}$ pairs, with $f$ corresponding to Levy, Pareto or Cauchy distribution,  the GCLT  predicts $\lambda_{min}$ to be distributed as $f(\lambda_{min};  \tilde{a},\tilde{b}, \tilde{c},\tilde{\delta})$. here $\tilde{a}=\tilde{a_0}=1/2, \tilde{b}=\tilde{b_0}=1, \tilde{c}=\tilde{c_0} \sqrt{N}$  and $\tilde{\delta}=N \tilde{\delta_0}$ for Levy,  $\tilde{a}=\tilde{a_0}, \tilde{c}=N \tilde{c_0}$ for Pareto, $\tilde{a}=\tilde{a_0}=1, \tilde{b}=\tilde{b_0}=0, \tilde{c} = N \tilde{c_0}$ and $\tilde{\delta}=N \tilde{\delta_0}$ for Cauchy.  
Substituting eqs(\ref{yal1}, \ref{yap1}, \ref{yac1}) with $y=-\lambda_{min}$ and $\beta \to {\beta \over L^p}$ in eq.(\ref{aaq4}), the condition for the lower limit can  be given as follows
\begin{eqnarray}
{\tilde X\over N}   \; \; & \le &   {- w_a }. \label{zxl} 
 \end{eqnarray}
with $\tilde{X} = X( \tilde{a},\tilde{b}, \tilde{c},\tilde{\delta}, \tilde{t})$ with $X$ given by eq.(\ref{zqwl2},\ref{zqwp2},\ref{zqwc2}). 
 Here again the above conditions can be rewritten in terms of $ \tilde{a_0},\tilde{b_0}, \tilde{c_0},\tilde{\delta_0}$. For example, for Levy case, eq.(\ref{zqwl1}) gives $- {\tilde{\delta_0} \over L^p}  + \sqrt{2 \tilde{c_0} \over \beta L^p}  \ge  -w_a$. Clearly a finite $w_a$ for Levy case can be defined even for limit $L \to 0$ if $\tilde{\delta_0} < 0$ and/ or $\beta$ is large.
 For Pareto case, eq.(\ref{zqwp1}) can be approximated as $ {\tilde{c_0} \over L^p}   \ge - w_a $ (neglecting the contribution from logarithmic terms); as $\tilde{c_0} > 0$, a finite $w_a$ can always be defined (e.g. $w_a =0$). For Cauchy case,  the bound becomes
$- {\tilde{\delta_0} \over L^p}   \ge - w_a$ which can easily be fulfilled e.g with $\tilde{\delta_0} < 0$.
The condition (\ref{zq3}) for the lower limit on free energy can then be fulfilled for a random  potential with long range spatial decay for all three types of distributions i.e Levy, Pareto as well as Cauchy. 

\vspace{0.2in}

{\bf Quenched, finite variance distributions} Following the same reasoning as in the annealed case with finite variance, both $\Lambda_0$ and $\lambda_{min}$ approach   Gaussian distributions,  in the large volume limit, with mean and variance as $(N_1 N_2 \mu_0,  N_1 N_2 \nu^2_0)$  and $(N u_0, N \sigma_0^2)$ respectively. The latter along with eq.(\ref{abq3}) and eq.(\ref{abq4}) now give the conditions for  $w_a, w_b$ as follows:

\begin{eqnarray}
 {\mu_0 \over R^p}  \le { w_b \over R^{d+\epsilon}}, \hspace{1in}
 {u_0 \over L^p}  \ge - w_a, \label{fg8}
\end{eqnarray}
 Clearly, here again, a finite $w_b$ exists if $\mu_0 < 0$ or $p > d$; note the latter case is analogous to the corresponding  condition for clean potentials.   But, as $u_0$ is finite, and, $L \to \infty$  in thermodynamic limit, the 2nd condition above can be fulfilled for an arbitrary $w_a >0$ and for an arbitrary $p$.

\vspace{0.2in}
 
 {\bf Quenched, power law distributions} Proceeding as in the annealed case i.e using  $f(\Lambda^{(st)}_{\eta \eta}; a_0, b_0, c_0,\delta_0)$ for $\Lambda^{(st)}_{\eta \eta}$ and $f(\lambda^{(st)}_{min}; \tilde{a_0},\tilde{b_0}, \tilde{c_0},\tilde{\delta_0})$  for $\lambda^{(st)}_{min}$ for all $s,t$-pairs but now using eqs.(\ref{ydl1}, \ref{ydp1}, \ref{ydc1}), one can calculate $\langle \Lambda_0 \rangle$ and $\langle \lambda_{min} \rangle$. The latter along with eq.(\ref{abq3}) and eq.(\ref{abq4}) now give the conditions for  $w_a, w_b$ as follows:
 
  \begin{eqnarray}
{1 \over R^p} Y(a_0,c_0,\delta_0)   \; \; & \le & \; \;  { w_b \over R^{d+\epsilon}},   \label{ywb} \\
{1\over L^p } \; Y(\tilde{a_0},\tilde{c_0},\tilde{\delta_0}) \; \; & \ge & \; \; -w_a.    \label{ywa}
\end{eqnarray}
where $Y$ for the three cases is given by eqs.(\ref{ydl1}, \ref{ydp1}, \ref{ydc1}). 
As clear from the above,  eq.(\ref{ywb}) can be satisfied for arbitrary $p$ if $Y(a_0,c_0, \delta_0) < 0$ and, except for Cauchy case,  the latter can be achieved  even if $\delta_0 > 0$ (note $\delta_0$ corresponds to mean of the distribution $f(a_0, b_0, c_0, \delta_0)$ for case $a_0 >1$ which is expected to be positive for repulsive potential).
 
 Further as $Y(\tilde{a_0},\tilde{c_0},\tilde{\delta_0})$ is finite, left side of eq.(\ref{ywa}) approaches zero  for arbitrary $p >0$ in thermodynamic limit ($L \to \infty$) for all three distributions mentioned above. Any choice of $w_a >0$ therefore satisfies the condition (\ref{ywa}) and thereby indicates existence of the upper bound of free energy.

}

\begin{table}[ht]
  \begin{center}
\caption{{\bf \label{tab:table1}  Extensive limit of disordered many body interactions with arbitrary spatial dependence:}  The table describes the conditions on the distribution parameters,  {\it with arbitrary spatial dependence},  for which the ensemble averaged free energy is  extensive; (note columns 2nd, 3rd, 5th and 6th correspond to conditions (\ref{zxl0}), (\ref{zxu0}), (\ref{ydl}), (\ref{ydu}) respectively with $w_a, w_b=0$). 
 In presence of disorder, the spatial dependence of many body interactions is expected to manifest through the distribution parameters of the potentials. More specifically, the ensemble averaged mean of the potential in general has the same spatial decay rate as  its single replica. As intuitively clear,  the  presence of more parameters in a condition increases the probability to  fulfill it. For annealed case, the temperature also enters in the condition as a parameter, thereby helping the collective conspiracy of the parameters to achieve extensive limits. Here the symbol $G$ in $3^{rd}$ column refers to the geometric mean of a specific combination of parameters: $G_p=\prod_{k,l} \left[a_{kl} \;\left({\beta c_{kl}}\right)^{a_{kl}} \Gamma{\left(-a_{kl}, {\beta c_{kl}} \right)} \right]^{1\over N_1 N_2}$ for Pareto,  $G_c= {\rm e}^{-\beta \overline{\delta}_U} \; \prod_{k,l} \left[{\rm e}^{i \beta c_{kl}} \Gamma{\left(i \beta c_{kl}\right)} + {\rm e}^{- i \beta c_{kl} } \Gamma{\left(- i \beta c_{kl} \right)} \right]^{1\over N_1 N_2}$ for Cauchy case,  $G_s =  {\rm e}^{-\beta \overline{\delta}_U} \; \prod_{k,l} \left[{1\over \pi}  \; \sum_{n=1}^{\infty} {(-1)^{n+1} (\beta c_{kl})^{n a_{kl} } \over n!  }   \sin\left({n a_{kl} \pi \over 2}\right) \; \Gamma(n a_{kl}+1) \;  \Gamma \left(-n a_{kl}, {\beta (u-\delta_{kl})} \right) \right]^{1\over N_1 N_2}$ for Sym-stable case.
 Note in case of quenched Levy, annealed Cauchy and symmetric-stable (both annealed and quenched) distributions, the results mentioned in the table are applicable only  for the  restricted support (see text). The $4^{th}$ column   states whether the annealed disordered LRIs with specific distribution type given in column $1^{st}$ (and with $\mu, \nu^2, \overline{\delta}_U, \overline{c}_U \sim {1\over R^{\gamma}}$, $\gamma <d$)  can be extensive  i.e whether both the conditions in columns $2^{nd}$ and $3^{rd}$ can simultaneously be fulfilled for them (a brief explanation given in \cite{foot1}). The $7^{th}$ column contains the similar information for the quenched disorders (a brief explanation given in \cite{foot2}).  }
\begin{ruledtabular}
\begin{tabular}{|l|c|c|c|c|c|r|}
Distribution  & Annealed   & Annealed &  Annealed & Quenched   &  Quenched & Quenched  \\
    Type   & lower bound & upper bound & LRIs & lower bound & upper bound& LRIs\\
\hline
  Gaussian    & finite $u, \sigma^2, T$  &  $2 \mu k T  \le  \nu^2$   & yes\cite{foot1} & finite $u$   & $\mu <0$   & no \cite{foot2} \\
   Levy    &  finite $\overline{\delta}_L$   &  ${\overline\delta}_U +  \sqrt{2 k T  \; {\overline c}_U }  \le 0  $  & yes \cite{foot1}    &  $  \overline{\delta}_L <  \overline{c}_L$   &   $  \overline{\delta}_U \geq  \overline{c}_U$  & yes  \cite{foot2} \\
   Pareto      &   finite $c_k/T$   &  $ G_p \ge 1$  &  yes \cite{foot1}&
   $\sum_{k} {\alpha_{k} c_{k} \over 1-\alpha_{k}} \leq 0$     &   $\sum_{k,l} {\alpha_{kl} c_{kl} \over 1-\alpha_{kl}}  \geq 0$ & no\cite{foot2}
\\
   Cauchy      &  finite $\delta_k, c_k/T$    & $G_c \ge 2$        &  yes\cite{foot1} & $\overline{\delta}_L \le 0$     & $\overline{\delta}_U \ge 0$     & yes\cite{foot2}   \\
   Sym-St      &  finite $\delta_k, c_k/T$    &  $G_s \ge 1 $      & yes \cite{foot1}
   & $Y_L \le 0 $   & $Y_U \ge 0$ & may be\cite{foot2}  \\
\end{tabular}
\end{ruledtabular}
\end{center}
\end{table}

\begin{sidewaystable}[h!]
  \begin{center}
  \caption{{\bf \label{tab:table2}  Extensive limit, annealed disorder and low temperature:}  Besides spatial decay rate of the potential, the bounds on the averages in the annealed case can in general depend on temperature too. Here the 2nd and 3rd column describe the low temperature limit of the bounds on the distribution parameters for the ensemble averaged free energy to be extensive.  Assuming the distribution parameters (i.e $\sigma^2, \overline{\delta}_U,  \overline{c}_U$) appearing in column 3 with a spatial dependence of $1/r^{\gamma}$ type, 
 the columns 3rd and 4th predict if a $d$-dimensional system with specific disordered potential type is extensive. Note the case $\gamma >d$  for each disorder type  is similar to the clean case.}   
\begin{ruledtabular}
\begin{tabular}{|l|c|c|c|r|}
Distribution  & lower  & upper & Case ${1\over r^\gamma}$  & Case  ${1\over r^\gamma}$ \\
 Type      & bound & bound  & with $\gamma <d$ & with $\gamma \ge d$ \\
\hline
  Gaussian    & $\sigma^2 \le 0$ &  $  \nu^2 \ge 0$   & extensive if $\sigma^2=0$ & extensive if $\sigma^2=0$ \\ 
   Levy    &   $\overline{\delta}_L \ge - w_a$   &  
   $\overline{\delta}_U \le  {w_b\over R^{d+\epsilon}}$       &  extensive if $\overline{\delta}_U \le 0$, $\overline{\delta}_L$ finite  & extensive if $\overline{\delta}_L$ finite \\
   Pareto      &   $ \overline{c}_L \ge -w_a$   &$ \overline{c}_U \le  {w_b\over R^{d+\epsilon}}$  & not extensive (as $\overline{c}_U \ge 0$)  & extensive    
\\
   Cauchy      &  $\overline{\delta}_L +{\pi \over 2} \; \overline{c}_L \ge -w_a$   & $\overline{\delta}_U +{\pi \over 2} \; \overline{c}_U \le  {w_b\over R^{d+\epsilon}}$    &  extensive if $\overline{\delta}_U \le 0$,  $|\overline{\delta}_U| \ge {\pi \over 2}\; \overline{c}_U$, $\overline{\delta}_L$ finite & extensive if $\overline{\delta}_L$ finite\\
   Sym-St      &  $\overline{\delta}_L \ge -w_a$    &  $\overline{\delta}_U   \le  {w_b\over R^{d+\epsilon}}$     &   extensive if $\overline{\delta}_U \le 0$, $\overline{\delta}_L$ finite  & extensive if $\overline{\delta}_L$ finite\\
\end{tabular}
\end{ruledtabular}
\end{center}
\end{sidewaystable}

\end{document}